\documentclass[aps, prl, twocolumn, longbibliography]{revtex4-1}

\usepackage{palatino}
\usepackage[utf8]{inputenc}
\usepackage{epsf}
\usepackage{amsmath,amssymb}
\usepackage{calc}

\usepackage{float}

\usepackage[usenames,dvipsnames]{xcolor}

\usepackage{graphicx}
\usepackage{hyperref}
\usepackage{ulem}

\usepackage{cancel}

\newcommand{\ben}{\begin{equation}}
\newcommand{\een}{\end{equation}}
\newcommand{\bea}{\begin{eqnarray}}
\newcommand{\eea}{\end{eqnarray}}

\def\ket#1{\vert#1\rangle}

\def\sss{\scriptscriptstyle\rm}

\def\1s{_{1,\sss S}}
\def\2s{_{2,\sss S}}

\def\br{{\bf r}}
\def\bR{{\bf R}}

\def\bff{{\bf f}}
\def\bQ{{\bf Q}}

\def\dulr{{\underline{\underline{\bf r}}}}
\def\dulR{{\underline{\underline{\bf R}}}}

\DeclareMathAlphabet\mathbfcal{OMS}{cmsy}{b}{n}
\def\pq{{\mathbfcal{Q}}}

\begin{document}

\title{Electronic Coherences in Molecules: The Projected Nuclear Quantum Momentum as Hidden Agent}
%The Essential Role of the Nuclear Quantum Momentum for Electronic Coherences in Molecules
%title instead? Capturing electronic coherence in mixed quantum-lassical simulations of molecular dynamics
\author{Evaristo Villaseco Arribas}
\affiliation{Department of Physics, Rutgers University, Newark 07102, New Jersey USA}
\author{Neepa T. Maitra}
\affiliation{Department of Physics, Rutgers University, Newark 07102, New Jersey USA}
\date{\today}

\begin{abstract}
Electronic coherences are key to understanding and controlling photo-induced molecular transformations. We identify a crucial quantum-mechanical feature of electron-nuclear correlation, the projected nuclear quantum momenta, essential to capture the correct coherence behavior. For simulations, we show that, unlike traditional trajectory-based schemes, exact-factorization-based methods approximate these correlation terms, and correctly capture electronic coherences in a range of situations, including their spatial dependence, an important aspect that influences  subsequent electron dynamics and that is becoming accessible in more experiments. 
%Electronic coherences in molecules are key to understanding and controlling photo-induced chemical transformations, and can be directly studied in recent experiments. The traditional mixed quantum-classical methods used to computationally simulate these processes lack a crucial feature of electron-nuclear correlation, the nuclear quantum momentum, that leads to their erroneous predictions of the dynamics. Instead, we show that  recently-developed methods based on the exact factorization approach correctly capture electronic coherences in a wide range of situations, including their spatial dependence, an aspect not only becoming accessible in more experiments but also influencing  subsequent electron dynamics.        
\end{abstract}

\maketitle
Quantum electronic coherences are key factors influencing many photo-induced molecular processes.
%determining the motion of electrons in molecules and their coupling to nuclear motion, 
 They serve as  control knobs in chemical transformations
~\cite{KMRW23b,GMPLMK22, MKCBW23, DGK18, DKW22, VBRM17} and possibly impact photosynthetic energy flow in biomolecules~\cite{DPCASTM17,MOSBSG18}, as well as quantum information science processes \cite{Wasielewski2020}. Aside from the practical interest in creating desired products, 
studying the generation and evolution of  coherences, including decay and revival, reveals fundamental properties of how correlations between electrons as well as their interplay with nuclei affect dynamics. 
%{\sout{Experiments can now track how coherences evolve in time  with spatial resolution as well~\cite{LGIHM20,CKRASGM21,GMPLMK22,MKCBW23}.}}

Electronic coherences are usually defined in the Born-Oppenheimer (BO) representation, but nevertheless can be extracted from experiment for the physical reason that  away from  regions of strong non-adiabatic  coupling (NAC), components of the nuclear wavefunction on different BO surfaces evolve independently. 
The spatially-resolved coherence is defined as~\cite{VBRM17,MKCBW23,VMA24}
\begin{equation}
\Gamma_{jk}(\dulR,t) = \chi_j^*(\dulR,t)\chi_k(\dulR,t)
\label{eq:Gamma}
\end{equation}
for $j\neq k $ (and populations $j=k$), 
where the $\chi_j(\dulR,t)$ are projected nuclear wavefunctions defined via expanding the molecular wavefunction in the BO basis: $\Psi(\dulR,\dulr,t) = \sum_j \chi_j(\dulR,t)\Phi_{\dulR,j}(\dulr)$, with $\Phi_{\dulR,j}$ eigenstates of the BO Hamiltonian $H_{\rm BO}\ket{\Phi_{\dulR,j}} = E_j^{\rm BO} \ket{\Phi_{\dulR,j}}$,  where variables $\dulr$, $\dulR$ denote all  electronic and nuclear coordinates, respectively. While recent advances in experimental techniques can measure the electronic coherences with sub-\AA~spatial resolution~\cite{LGIHM20,CKRASGM21,GMPLMK22,MKCBW23}, most experiments instead measure the spatially-averaged coherence~\cite{KMRW23b,GMPLMK22, MKCBW23, DGK18, DKW22, VBRM17},  $\Gamma_{jk}(t) = \int d\dulR \Gamma_{jk}(\dulR,t)$, often referred to as coherence {\it tout court}.

The electronic coherence  thus measures the overlap of the projected wavefunctions.  Semiclassical considerations have identified three mechanisms for how this evolves in time, in particular for decoherence (i.e. its decay)~\cite{FH03}: nuclear density overlap decay, pure dephasing, 
%(from different phases for different $\dulR$ giving cancellations in the spatial integral), 
and electronic state transitions. But open questions remain  
%when different surfaces have different slopes, in regions of negligible NAC, these wavefunctions evolve away from each other leading to decaying coherence, a phenomenon often referred to as decoherence. {\color{magenta}\it i think the next paragraphs all the way to the next page actually need to be heavily editted...we have to be clearer in what we are trying to say and relate to these concepts also} \textcolor{blue}{ While three main mechanisms of decoherence have been identified~\cite{FH03,GBV20} to be nuclear overlap decay, dephasing (``phase jitter"), and electronic state transitions, with time-evolution determined either by semiclassical ans\"atze under some assumed behavior in certain limits,  or through a full quantum-dynamical calculation, 
%providing new insights into the interplay of the three identified mechanisms~\cite{VBRM17},
%a complete and general picture is missing. 
%A full quantum-dynamical calculation provided new insight into the interplay of the three identified mechanisms~, but 
%..or are there other terms that play a key role in certain cases?
in the nature of the electron-nuclear correlation in influencing coherences, decoherences, and recoherences (i.e. revivals after decoherence):  The semiclassical analyses focus on the spatially-averaged coherence and hold under some assumed wavepacket form and behavior in certain limits~\cite{GBV20}, so then do these three mechanisms provide a complete picture fully accounting for correlated electron-nuclear dynamics, or do other factors play a key role? 
While the three mechanisms, and their interplay, have been identified in quantum dynamical simulations~\cite{VBRM17}, how do they explicitly appear in the  equations of motion? Computational challenges in modeling complex molecules mean that mixed quantum-classical (MQC) methods are  mostly used, where one propagates an ensemble of classical nuclear trajectories each carrying a set of quantum electronic coefficients determined self-consistently through a prescribed quantum-classical feedback; can we precisely identify which quantum properties of the nuclear system influence the evolution of electronic coherence and need to be approximated in such approaches? 
%{\sout{Without assumptions made in the thawed gaussian wavepacket approach~\cite{GBV20} and including non-adiabatic couplings,}}
The most  
commonly-used methods, Ehrenfest and surface-hopping~\cite{T90,T98} are both fundamentally unable to correctly capture coherence, with overcoherence in Ehrenfest and internal inconsistency in surface-hopping
%while the electronic coefficients in Ehrenfest always remain coherent, those in surface-hopping are inconsistent with the trajectory-evolution and 
%and use of  a largely  {\it ad hoc} decoherence procedure
~\cite{T90,WAP16,CB18,SJLP16}.  
Ref.~\cite{EL20} pointed out that the commonly-used, largely  {\it ad hoc}, decoherence corrections are fundamentally flawed since they act in a state-wise manner while coherence is a state pair-wise property.
Further, if an experiment cannot resolve the  spatial character in Eq.~\ref{eq:Gamma}, an inherent signature of correlation, does it matter if the MQC method does not capture it well if it captures the spatial-average well?

Here we show that even when only the spatially-averaged coherence is measured, the underlying spatial-dependence strongly influences the time-dependence  (beyond pure dephasing), and that the projected nuclear quantum momenta, $\nabla_\nu\vert\chi_j\vert/\vert\chi_j\vert$, are key to capture the correct behavior. Thus electron-nuclear correlation terms that go beyond  the earlier semiclassical analyses are essential.  
Even in cases where traditional trajectory-based MQC simulations yield the correct coherence over the duration of one interaction event, their wrong spatially-resolved coherence leads to poor behavior at later times. Instead, methods based on the exact factorization (EF) approach~\cite{MAG15,AMAG16,HLM18,VM23,DRM24} better approximate these terms and therefore the spatial structure, giving greatly improved predictions, distinguishing between coherence of wavefunctions on parallel surfaces (unlike {\it ad hoc} decoherence-corrected methods) and gradual decoherence of non-parallel ones (unlike Ehrenfest). While existing EF-based approximations contain a crucial dependence on the overall nuclear quantum momentum, we identify their neglect of the individually projected quantities as the culprit for not accurately capturing recoherence events in regions of negligible NAC.

Coherences evolve due to population transitions from NACs ($\langle\Phi_{\dulR,j}\vert\nabla_\nu \Phi_{\dulR,k}\rangle$ and $\langle\Phi_{\dulR,j}\vert\nabla_\nu^2 \Phi_{\dulR,k}\rangle$),  but also away from those regions when more than one BO surface is populated. To avoid conflating effects from NACs, we begin by considering the exact equation of motion for the spatially-resolved coherence in a situation where all NACs are zero:
\begin{eqnarray}
i \partial_t \Gamma_{jk}(\dulR,t) &=& \Delta E_{kj}(\dulR)\Gamma_{jk}(\dulR,t)+\sum_\nu \frac{1}{2M_\nu}\bigg(\chi_k(\dulR,t)
\nonumber \\
&&\nabla_\nu^2\chi_j^* (\dulR,t) -\chi_j^*(\dulR,t)\nabla_\nu^2\chi_k(\dulR,t)\bigg)
\label{eq:dcohRt/dt}
\end{eqnarray}
where $\Delta E_{kj}(\dulR) = E^{\rm BO}_k(\dulR) - E^{\rm BO}_j(\dulR)$ and the sum over $\nu$ is a sum over all nuclei. Eq. (S.8) in the SM gives the equation including the NACs. Atomic units ($\hbar=m_e=e^2=1$) are used throughout this article. The first term on the right of Eq.~\ref{eq:dcohRt/dt} can be absorbed in a phase, while the evolution of the coherence magnitude depends on the instantaneous curvatures of the BO projected wavefunctions.
%such that the magnitude of the spatially-resolved coherence depends only the curvatures of the BO projected wavefunctions and not explicitly on the shape of the BO surfaces. 
Instead, when integrated over space, the explicit dependence on these curvatures vanishes, and we find
\ben
i\partial_t \Gamma_{jk}(t) = \int \Delta E_{kj}(\dulR) \Gamma_{kj}(\dulR,t) d\dulR
\label{eq:dcoh/dt}\,,
\een
that is,  the spatially-averaged coherence explicitly depends on the instantaneous  relative difference in shape of the BO surfaces weighted by the spatially-resolved coherence.
%For the magnitude of the coherences, we obtain  (away from any NAC)
%\bea
%\nonumber
%\partial_t \vert \Gamma_{jk}(\dulR,t) \vert^2 &=& {\rm Im}\sum_\nu \frac{-1}{M_\nu}
%\left(\vert\chi_j(\dulR,t)\vert^2 \chi_k^*(\dulR,t)\nabla_\nu^2\chi_k(\dulR,t)\right.\\
%&+& \left.\vert\chi_k(\dulR,t)\vert^2 \chi_j^*(\dulR,t)\nabla_\nu^2\chi_j(\dulR,t)\right)\,
%\eea
%which is dependent on the curvature of the BO projected wavefunctions but not explicitly dependent of the shape of the BO surfaces, while the spatially-integrated coherence evolves as 
%\ben
%\partial_t \vert \Gamma_{jk}(t) \vert^2 =2 {\rm Im}\left( \Gamma_{kj}(t) \int \Delta E_{kj}(\dulR)\Gamma_{jk}(\dulR,t) d\dulR\right),
%\label{eq:dmagcoh/dt}
%\een
%which instead, is explicitly dependent on the relative shape of the BO surfaces and not explicitly dependent of the curvature of the projected wavefunctions. 
We make two key observations from Eqs.~\ref{eq:dcohRt/dt}--~\ref{eq:dcoh/dt}:  First,  \textit{without correct spatial dependence of the coherence (a dependence that inherently signifies electron-nuclear correlation), the time-evolution of the spatially-averaged coherence or populations will be wrong}. Second, \textit{for parallel surfaces (no pure dephasing) the magnitude of the spatially-averaged coherence is constant  ($\Gamma_{jk}(t) = e^{i(E_k - E_j) t}\Gamma_{jk}(0)$), but there is a spatial structure to these quantities that does evolve in time (last two terms of Eq.~(\ref{eq:dcohRt/dt})).}

%While nuclear motion  influences electronic coherences 
% A deeper understanding of how nuclear motion influences coherences requires to discern effects arising from classical point-like nuclear motion within an ensemble and quantum effects from nuclear wavepacket delocalization. 
 A deeper understanding of how nuclear motion influences coherences requires to discern local electron-nuclear correlation effects arising from a classical-like treatment of the nuclear motion via an ensemble of trajectories from non-local effects from nuclear wavepacket delocalization.
 To address this, we turn to the EF,  which, unlike Ehrenfest and surface-hopping, allows a formulation of trajectory-based equations defining exact unique forces on the nuclear trajectories when they are treated classically~\cite{LRG22,AAG14,AAG14b}, and evolution equations for the populations and coherences. In EF, the  molecular wavefunction is represented by a single correlated product, $\Psi(\dulr,\dulR,t) = \chi(\dulR,t)\Phi_\dulR(\dulr,t)$ with $\int \vert\Phi_\dulR(\dulr,t)\vert^2 d\dulr =1\,\forall\,\dulR,t$~\cite{Hunter75,Hunter_IJQC1975_2,Hunter_IJQC1980,H81,Hunter_IJQC1982,GG14,AMG10,AMG12,AG21,VAM22}. (See also Supplementary Material (SM) for brief details from these works~{\footnote{{See Supplemental Material which includes Refs.~\cite{Hunter75,Hunter_IJQC1975_2,Hunter_IJQC1980,H81,Hunter_IJQC1982,GG14,AMG10,AMG12,AG21,VAM22,AAG14,AAG14b,MAG15,ATC18,LRG22,WRB04,BM98,BQM00,S09,MSFS17,MGMS14,GCTMQC,PyUNIxMD,VMA24} }}}). The EF yields the notion of a unique 
nuclear wavefunction  that satisfies a Hamiltonian evolution and whose modulus and phase give the exact nuclear density and nuclear current-density, ${\bf J}_\nu(\dulR,t)$, of the molecular wavefunction~\cite{AMG10,AMG12}, central concepts to set up an exact trajectory-based approach.
%This allows a formulation of {\it exact} unique trajectory-based equations defining exact unique forces on the nuclear trajectories when they are treated classically~\cite{LRG22,AAG14,AAG14b}, and evolution equations for the populations and coherences, as we will see shortly.

Writing the EF nuclear wavefunction in terms of an amplitude  and phase, $\chi(\dulR,t)=e^{iS(\dulR,t)}\vert\chi(\dulR,t)\vert$, and likewise for the projected BO wavefunctions, $\chi_k(\dulR,t) = \vert \chi_k(\dulR,t)\vert e^{iS_k(\dulR,t)}$, we have, in the limit of negligible NAC, (see Eq. (S.18) in the SM for the general case)
\begin{eqnarray}
&&\partial_t \vert\Gamma_{jk}(\dulR,t)\vert =-\vert\Gamma_{jk}(\dulR,t)\vert\sum_\nu \Bigg\{ \frac{\big(\nabla_\nu-2\overline{\mathfrak{S}}_{\nu,jk}\big)\cdot\bold{J}_\nu}{\vert\chi\vert^2} \nonumber \\
&&+\sum_n\frac{\vert C_n\vert^2}{2M_\nu}\bigg[\overline{\bigg(4\pq_{\nu,jk}-2\nabla_\nu\bigg)\cdot\big(\mathbf{f}_{\nu,n}-\mathbf{f}_{\nu,jk}\big)} +4\mathfrak{S}_{\nu,n}\cdot\mathbf{f}_{\nu,n}\bigg]\Bigg\} \nonumber \\
\label{eq:exactdcohdt2}
\end{eqnarray}
In Eq.~\ref{eq:exactdcohdt2}, all quantities on the right are functions of $\dulR$ and $t$, an overline indicates the average over $j,k$ ($\overline{g}_{jk} = (g_j+ g_k)/2$), and we have defined $\bff_{\nu,k} = \nabla_\nu(S_k - S)$.    Further, we defined the nuclear quantum momentum $\mathbf{Q}_{\nu}$ and projected components $\mathbfcal{Q}_{k,\nu}$ for each state $k$  through 
\begin{equation}
{\bf Q}_\nu=-\frac{\nabla_\nu \vert\chi(\dulR,t)\vert}{\vert\chi(\dulR,t)\vert} \; {\rm and} \; \mathbfcal{Q}_{k,\nu}(\dulR,t)= -\frac{\nabla_\nu \vert\chi_k(\dulR,t)\vert}{\vert\chi_k(\dulR,t)\vert}
\end{equation}
while $\mathfrak{S}_{\nu,k}$ is the ``reduced" contribution
\begin{equation}
\mathfrak{S}_{\nu,k}(\dulR,t)=-\frac{\nabla_\nu \vert C_k(\dulR,t)\vert}{\vert C_k(\dulR,t)\vert} = {\mathbfcal Q}_{\nu,k} - {\bf Q}_\nu
\end{equation}
with $C_k(\dulR,t) = \chi_k(\dulR,t)/\chi(\dulR,t)$, which represent coefficients of the conditional electronic wavefunction $\Phi_\dulR(\dulr,t)$ when expanded in the BO basis, $\Phi_\dulR(\dulr,t) = \sum_n C_n(\dulR,t)\Phi_{\dulR,n}(\dulr)$.

Our third key observation follows from Eq.~(\ref{eq:exactdcohdt2}):
While the first term on the right represents a convective contribution to the time-derivative (see more shortly),  \textit{both the projected quantum momentum ${\mathbfcal{Q}}_{\nu,k}$ and the reduced contribution $\mathfrak{S}_{\nu,k}$, play a crucial role in capturing accurate spatially-resolved populations and coherences, when away from NAC regions}. The only other term driving (the magnitude of) the coherence or population evolution, $\sum_n |C_n|^2\nabla_\nu\cdot(\bff_{\nu,k}+\bff_{\nu,j}  - 2\bff_{\nu,n}) = \sum_n |C_n|^2\nabla^2_\nu(S_k + S_j - 2S_n)$, depends on differences in curvature of the adiabatic phases, which semiclassically relates to the curvatures of the BO surfaces (see more shortly). A fourth key observation is that \textit{in regions where the coherence between two states has locally collapsed to zero, only the terms depending on $\mathfrak{S}_{\nu,j(k)}$ survive to drive time-evolution of the spatially-resolved coherences and populations; thus, this is responsible for recoherence effects away from NACs,  and will be discussed in more detail shortly.}
%\begin{widetext}
%\ben
%\partial_t \vert\chi_k(\dulR,t)\vert^2 =  -\vert\chi_k(\dulR,t)\vert^2 \sum_\nu \left[ \frac{\nabla_\nu\cdot {\bf J}_\nu - \mathfrak{S}_{\nu,k}\cdot{\bf {J}_\nu}}{\vert\chi\vert^2}
%+\frac{1}{M_\nu}\left(-2 {\bQ}_{k,\nu}\cdot\left(\bff_k(1 - |C_k|^2) - \sum_{n\neq k} \vert C_n\vert^2 \bff_n\right)
%+\left.\nabla_\nu\bff_k(1 - \vert C_k\vert^2) - \sum_{n\neq k} %\vert C_n\vert^2 \nabla_\nu\cdot \bff_n + 2\sum_n \vert C_n\vert^2\mathfrak{S}_{\nu,n}\cdot \bff_n\right)
%\right]
%\een
%\end{widetext}

%Next we turn to how the coherence evolves in trajectory-based MQC methods that 

%Because it is rooted in the EF picture, which enables a 
 The unique and unambiguous definition of the total nuclear density and current-density rooted in the EF, allows us to derive an {\it exact trajectory-based} equation 
 %for the populations and coherences 
 from
Eq.~(\ref{eq:exactdcohdt2}).   We
represent the nuclear density as a sum over $\delta$-functions (or very narrow Gaussians) centered at a trajectory position $\dulR^{(I)}(t)$
\ben
\vert \chi(\dulR,t)\vert^2 \rightarrow \frac{1}{N_{tr}}\sum_I\delta(\dulR - \dulR^{(I)}(t))
\een
where $\dulR^{(I)}(t)$ satisfy classical Newton's equations~\cite{MAG15,AMAG16} with a generalized Lorentz force dependent on $\Phi_\dulR$~\cite{AAG14,AAG14b,LRG22}. %Eq.~\ref{eq:grad_phase} 
The  gradient of the phase $S$ becomes $\nabla_\nu S \rightarrow M_\nu \dot{\dulR}(t) - {\bf A}$ where ${\bf A}$ is the vector potential in the nuclear Hamiltonian (see SM), and the time-derivative along the trajectory given by the convective derivative $d/dt = \partial_t + \sum_\nu\dot{\bR}^{(I)}_\nu\cdot\nabla_\nu$. 
The spatially-resolved and -averaged coherences and populations 
%of Eqs.~(\ref{eq:Gamma})--~\ref{eq:Gamm} 
become trajectory-ensembles:
\bea
\nonumber
\Gamma_{jk}(\dulR,t) &\rightarrow& \frac{1}{N_{tr}}\sum_I\delta(\dulR - \dulR^{(I)}(t))C_j^{(I)*}(t)C_k^{(I)}(t)\\
%\een
%and spatial-averaging becomes a trajectory ensemble:
%\ben
\Gamma_{jk}(t) &\rightarrow& \sum_I C_j^{(I)*}(t)C_k^{(I)}(t)
\eea
Then Eq.~\ref{eq:exactdcohdt2} becomes 
\begin{eqnarray}
&&\frac{d\vert C_j^*C_k\vert}{dt}^{(I)} =-\left\vert C_j^{*(I)}C_k^{(I)}\right\vert \,\sum_{\nu,n} \frac{\vert C_n^{(I)}\vert^2}{2M_\nu}\nonumber \\
&&\bigg[\overline{\left(4\pq_{\nu,jk}^{(I)}-2\nabla_\nu\right)\cdot\left(\mathbf{f}_{\nu,n}^{(I)}-\mathbf{f}_{\nu,jk}^{(I)}\right)}+4\mathfrak{S}_{\nu,n}^{(I)}\cdot\mathbf{f}_{\nu,n}^{(I)}\bigg]
\label{eq:dCsqdttraj}
\end{eqnarray}
%\bea
%\nonumber
%\frac{d}{dt}(C_j^{*(I)}(t)C_k^{(I)}(t)) = - C_j^{*(I)}C_k^{(I)}\sum_\nu \sum_n \vert C_n^{(I)}\vert^2 .\\
%\left[\left(\frac{2\mathbfcal{Q}^{(I)}_{\nu,k} +\nabla_\nu}{M_\nu}\right)\cdot(\bff^{(I)}_{\nu,k} - \bff^{(I)}_{\nu,n}) + 2\frac{\mathfrak{S}^{(I)}_{\nu,n}}{M_\nu}\cdot \bff^{(I)}_{\nu, n}\right]
%\label{eq:dCsqdttraj}
%\eea
%\textcolor{blue}{Note this equation can be rewritten as (actually from this equations everything is much more clear as we separate out the contribution of the QMOM which CTMQC computes and the contribution from crunch which CTMQC misses. Also you can clearly see the crunch term is only the crunch term for the current state $k$ and is multiplied only by $\vert C_k\vert^2$ and not by $\vert C_n\vert^2\vert C_k\vert^2$ and does not involve a sum on states): {\color{purple} sure we can do that but write it for CjCk, and what happened to the div.f terms?}
%\begin{eqnarray}
%\frac{d\vert C_k^{(I)}(t)\vert^2}{dt} &=& -\vert C_k^{(I)}\vert^2\frac{\mathfrak{S}_{\nu,k}^{(I)}}{M_\nu}\cdot\mathbf{f}_{\nu,k}^{(I)}+\vert C_k^{(I)}\vert^2\sum_\nu\frac{2\mathbf{Q}^{(I)}_{\nu}}{M_\nu}\cdot \nonumber \\
%&&\sum_n\vert C_n^{(I)}\vert^2\Delta\mathbf{f}_{\nu,nk}^{(I)} \nonumber \\
%&=& \frac{\nabla_\nu\vert C_k^{(I)}\vert^2}{M_\nu}\cdot\mathbf{f}_{\nu,k}^{(I)}+\vert C_k^{(I)}\vert^2\sum_\nu\frac{2\mathbf{Q}^{(I)}_{\nu}}{M_\nu}\cdot \nonumber \\
%&&\sum_n\vert C_n^{(I)}\vert^2\Delta\mathbf{f}_{\nu,nk}^{(I)}
%\end{eqnarray}
%}
where all are functions of $t$ alone, and we use the short-hand $g^{(I)} = g(\dulR^{(I)}(t))$.

Away from any NAC, Eq.~\ref{eq:dCsqdttraj} gives the exact equation for the magnitude of the electronic populations and coherences that trajectory-based methods should be aiming for. The full equation including the NACs, and phases, is given in the SM (Eq.~(S.33--S.35)). The $I$-dependence gives the spatially-resolved character in the trajectory-picture:  while the electronic populations ($C_j^{*(I)}C_j^{(I)}(t)$) and coherences ($C_j^{*(I)}C_{k\neq j}^{(I)}(t)$) for each individual trajectory evolve in time in the case of parallel surfaces, their sum over trajectories should remain constant in the limit of a large sampling of the initial distribution, analogous to the discussion below Eq.~(\ref{eq:dcoh/dt}). Time-dependence of the individual $C_k^{(I)}(t)$ indicates spatial-dependence in the coefficients, crucial for getting the correct populations and coherences, even when only ensemble-averaged quantities are measurable, as discussed earlier and as we will see explicitly shortly. %The time-evolution of the magnitude of the coherences is obtained from Eq.~\ref{eq:dCsqdttraj} with Eq.~\ref{eq:dcohdt2}. 

The traditional trajectory-based methods (Ehrenfest, surface-hopping) give strictly zero time-evolution throughout the ensemble when NACs are negligible because they have no non-local quantum momentum terms, and {\it ad hoc} decoherence corrections to surface-hopping cause spurious decays. Our examples will demonstrate this has drastic consequences for intermediate and long-term coherence and population dynamics. 
The prime importance of the projected quantum momenta in Eq.~\ref{eq:dCsqdttraj} is, on the other hand, partially recognized in the EF-based CTMQC method~\cite{MAG15,AMAG16}, which approximates these terms.
Derived from a well-defined series of approximations on the exact electronic and nuclear equations, the resulting CTMQC equations neglect $\mathfrak{S}_{\nu,k}$, effectively approximating $\mathbfcal{Q}_{\nu,k}$ by $\bQ_\nu\,\,\,\forall\, k,\nu$. 
%The upshot is that in many cases, but not all, coherence is well-captured.
We find (again, for negligible NAC)
%{\color{purple} write for $C_jC_k$}
%\bea
%\nonumber\left(\frac{d\vert C_k^{(I)}(t)\vert^2}{dt}\right)^{\rm exact}=\left(\frac{d\vert C_k^{(I)}(t)\vert^2}{dt}\right)^{\rm CT}+\vert C_k^{(I)}\vert^2 \sum_{\nu,n}\vert C_n^{(I)}\vert^2 \\
%\nonumber\bigg[\bigg(\frac{\nabla_\nu-2\mathfrak{S}_{\nu,k}^{(I)}}{M_\nu}\bigg) \cdot(\bff^{(I)}_{\nu,k} - \bff^{(I)}_{\nu,n})+2\frac{\mathfrak{S}^{(I)}_{\nu,n}}{M_\nu}\cdot\bff^{(I)}_{\nu,n} \\
%+\frac{2\mathbfcal{Q}_{\nu,k}^{(I)}}{M_\nu}\cdot(\delta\mathfrak{f}_{\nu,k}^{(I)} - \delta\mathfrak{f}_{\nu,n}^{(I)})\bigg]\,\,\,\,\,\,\,\,\,\,\,\,\,\,\,\,\,\,\,\,
%\eea
%\begin{eqnarray}
%\left(\frac{d(\vert C_j^*C_k\vert)}{dt}\right)^{(I)}_{\mathrm{exact}} &=&\left(\frac{d(\vert C_j^*C_k\vert)}{dt}\right)^{(I)}_{\mathrm{CT}}+(\vert C_j^{*}C_k\vert)^{(I)} \Bigg\{\sum_\nu \frac{1}{2M_\nu}\nonumber \\
%&&\sum_l\vert C_l^{(I)}\vert^2\bigg[\overline{\left(4\mathfrak{S}_{\nu,jk}^{(I)}-\nabla_\nu\right)\cdot\Delta\mathbf{f}_{\nu,ljk}^{(I)}}\nonumber \\
%&&+4\mathfrak{S}_{\nu,l}^{(I)}\mathbf{f}_{\nu,l}^{(I)}\bigg]
%\label{eq:cohct}
%\end{eqnarray}
%Evaristo, tried again to put eqn in two lines but kept your original tex above. The eqn number is squished again but maybe can make it work...
\begin{eqnarray}
\nonumber 
\left(\frac{d\vert C_j^*C_k\vert}{dt}\right)^{(I)}_{\mathrm{exact}} =\left(\frac{d\vert C_j^*C_k\vert}{dt}\right)^{(I)}_{\mathrm{CTMQC}}+\vert C_j^{*(I)}C_k^{(I)}\vert \\
\sum_{\nu,n} \frac{\vert C_n^{(I)}\vert^2}{2M_\nu}\bigg[\overline{\big(4\mathfrak{S}_{\nu,jk}^{(I)}-2\nabla_\nu\big)\cdot\big(\mathbf{f}_{\nu,n}^{(I)}-\mathbf{f}_{\nu,jk}^{(I)}}\big)+4\mathfrak{S}_{\nu,n}^{(I)}\cdot\mathbf{f}_{\nu,n}^{(I)}\bigg]\nonumber\\
\label{eq:cohct}
\end{eqnarray}
This expression assumes the semiclassical approximation for the  gradient of the phase of the electronic coefficients $\bff_{\nu,l}^{(I)}=-\int_{0}^t\nabla_\nu E_l^{(I)}\,dt$, valid away from NACs~\cite{AMAG16}. 
%+\delta\mathfrak{f}_{\nu,l}^{(I)}$ is approximated by its semiclassical approximation neglecting $\delta\mathfrak{f}_{\nu,l}^{(I)}$.

A consequence of CTMQC's neglect of $\mathfrak{S}$ would be spurious population transfer in regions of zero NAC that is however fixed in implementations by redefining the quantum momentum to enforce no net transfer in these cases~\cite{MAG15,AMAG16,MATG17}. However, another consequence, that is not fixed by this redefinition, is CTMQC's inability to capture recoherence away from NAC regions as noted in the fourth observation made earlier. Consider 
%the exact trajectory-equation of motion for the coherences, Eq.~(\ref{eq:dCsqdttraj}), in 
the situation where, at some time $t$, $C_k^{(I)}(t) = 0$ for trajectory $I$ and  $C_j^{(J)}(t) = 0$ for trajectory $J$. 
If the forces on the nuclei cause the two trajectories to move to the same spatial region, the magnitude of the coherence $\Gamma_{jk}(t)$ should increase from zero for both these trajectories. Now, in the exact trajectory equation  Eq.~(\ref{eq:dCsqdttraj}) for trajectory $I$, the $|C_k^{(I)}(t)|$  in front of the sum on the right sets all terms to zero except for the term involving  $\pq_{\nu,k}^{(I)}$ since in that term $|C_k^{(I)}(t)|$ is in the denominator:
\ben
\frac{d}{dt}\vert C_j^*C_k\vert^{(I)} = \vert C_j^{(I)}\vert \sum_\nu\sum_n \frac{2 |C_n^{(I)}|^2}{M_\nu} \nabla_\nu|C_k^{(I)}|\cdot(\bff_{\nu,n}^{(I)} - \bff_{\nu,k}^{(I)})
\label{eq:recoherence}
\een
This is non-zero when for a neighboring trajectory $J$, %has a non-zero coefficient 
$C_k^{(J)}(t) \neq 0$, (i.e. $\nabla_\nu |C_k| \neq 0$). This spatial-dependence of the coefficients, contained in the reduced projected quantum momentum $\mathfrak{S}_{k,\nu}$, is key to the growth of the coherence. CTMQC neglects this term and as a result, CTMQC fails to capture recoherence. 

%\textcolor{blue}{Note also that when considering the time-evolution of the electronic populations in regions of zero NACs $\sum_I(d\vert C_k^{(I)}\vert^2/dt)_{\mathrm{exact}}=\sum_I(d\vert C_k^{(I)}\vert^2/dt)_{\mathrm{CT}}+\sum_I\vert C_k^{(I)}\vert^2\sum_{\nu,n}\vert C_n^{(I)}\vert^2\mathfrak{S}_{\nu,n}^{(I)}\mathbf{f}_{\nu,n}^{(I)}=0$. And it is precisely the neglect of $\mathfrak{S}$ that CTMQC might suffer from spurious transfer if the quantum momentum is not redefined~\cite{MATG17,VAM22}}

\begin{figure}
    \centering
\includegraphics[width=0.5\textwidth]{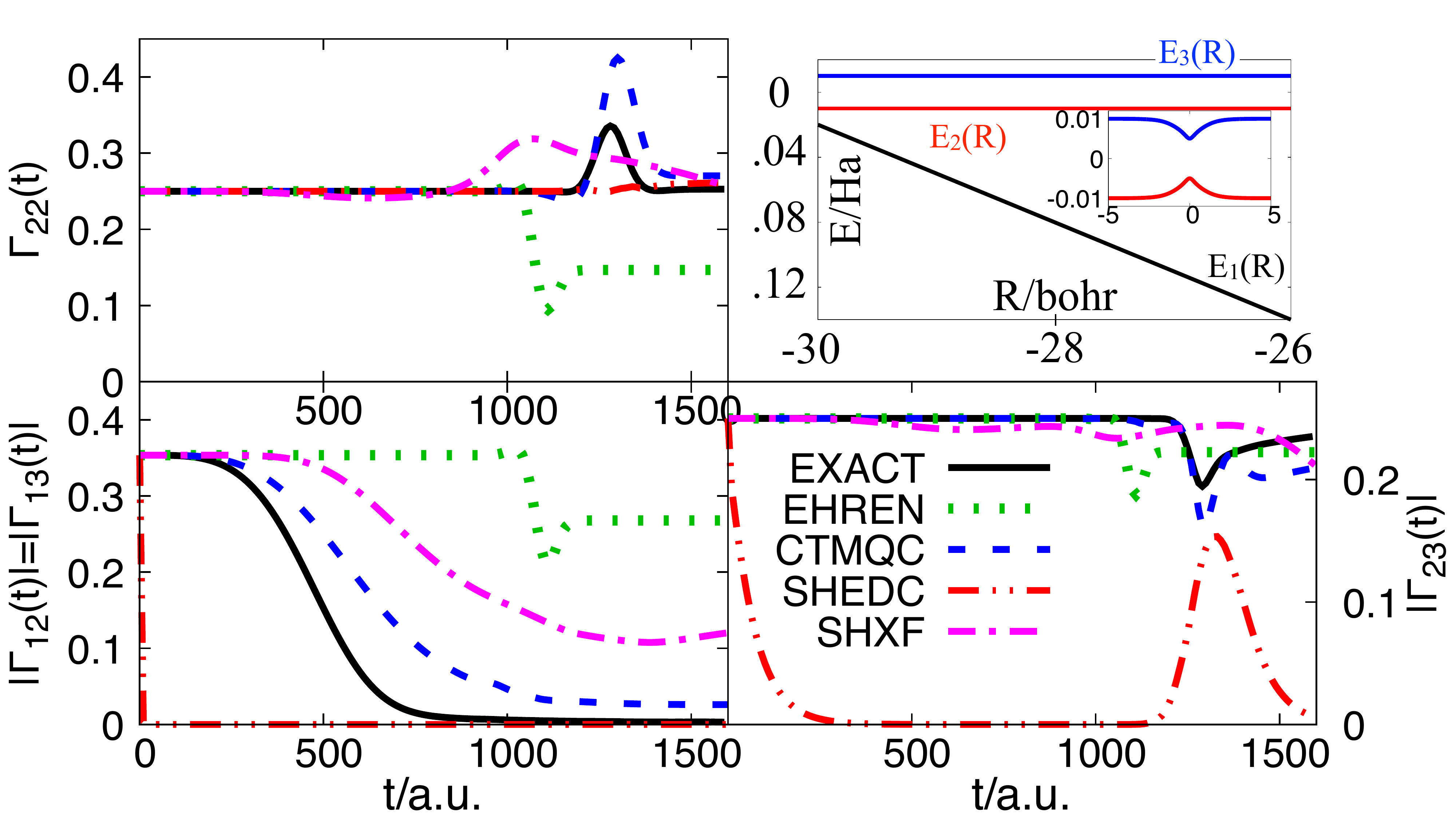}
    \caption{Electronic state populations of the first excited state (top left), adiabatic PES (top right), magnitude of electronic coherences between non-parallel { states 1 and 2 (or 3)} (bottom left) and parallel states  2 and 3 (bottom right) for EL20-SAC. The NAC is only non-zero near the SAC localized at $R = 0$.}
    \label{fig:coh-el20sac}
\end{figure}

Our first example, 
%is a slight variation on the one-dimensional three-state model of Ref~\cite{EL20}, 
EL20-SAC, consists of two parallel electronic states that eventually reach a single avoided crossing (SAC)~\cite{T90}, and a third non-parallel state uncoupled to the others~\cite{EL20} (inset of Fig.~\ref{fig:coh-el20sac}). The Hamiltonian is given in the SM. 

The system is initialized in a $50:25:25$ coherent superposition of three gaussian nuclear wavepackets with center momentum $40$ bohr$^{-1}$ and position $-26$ bohr.
%\begin{equation}
%\vert\Psi (R,0)\rangle=\chi(R,0)\left(\vert\phi_{R,1}\rangle+\vert\phi_{R,2}\rangle/2+\vert\phi_{R,3}\rangle/2    \right)
%\label{eq:superpos}
%\end{equation}
%with $\chi(R,0)=\left(\frac{2\pi^{1/2}}{3\sigma}\right)^{1/2}e^{-\frac{(R-R_0)^2}{2\sigma^2}+i(R-R_0)P_0}$ . The initial variance, position, momentum are $(\sigma,R_0,P_0)=(1\,\mathrm{bohr},-26\,\mathrm{bohr},40\,\hbar/\mathrm{bohr})$. 
%{\color{magenta}Evaristo, you're not going to like this but let me suggest it anyway -- what do you tyhnk of putting Eq 15 and the $\chi(R,0)$ eqn in the white space in the lower panel of figure? }
The model illustrates fundamental aspects of electronic coherences. First,  their pair-wise nature~\cite{EL20}: coherence between the pair of parallel states should be maintained while coherence between non-parallel states should be lost due to diverging wavepackets. Second, the need to accurately describe the spatially-resolved coherence, as this is key to correctly capture the dynamics even of spatially-averaged quantities, when later entering  into the NAC region. 
%We find that traditional methods fail miserably, while CTMQC does well. 

Figure~\ref{fig:coh-el20sac} shows the magnitude of the spatially-averaged electronic coherences~{\footnote{{Note that here we have shown $\int \vert \Gamma_{ij}(R,t) \vert dR$ as the magnitude of the coherence, rather than $\vert \int \Gamma_{ij}(R,t) dR \vert$, i.e. the measure we are showing does not include dephasing.} }} 
%between the parallel states (upper panel), non-parallel states (middle panel), 
and first-excited population,  obtained from Ehrenfest {(EHREN)}, surface-hopping with energy-based decoherence correction (SHEDC)~\cite{GP07,GPZ2010}, EF-based CTMQC,  EF-based surface-hopping (SHXF)~\cite{HLM18,PyUNIxMD}, and the exact calculations. Consider first  before the SAC is reached. As expected, Ehrenfest captures the spatially-averaged coherence between the parallel surfaces but fails to capture the decoherence between the non-parallel surfaces, while in SHEDC the parallel coherence erroneously decays and the non-parallel  decays much too quickly. Instead, CTMQC does a good job for both. The more approximate SHXF predicts too long  non-parallel coherence decay time, and the parallel coherence shows some deviation from constant. All methods capture the (trajectory-averaged) populations well at early times. However, once they reach the SAC, only CTMQC reasonably captures the population and coherence dynamics at the right time, albeit with an overestimation, before settling to about the right values. SHXF gets the trend in the right direction, but the timing and duration of the event is too long. The traditional  trajectory-based methods are completely wrong: SHEDC shows hardly any transfer and wrong coherence behavior, and Ehrenfest's population goes in the wrong direction. 
We note that the thawed Gaussian approach of Ref.~\cite{GBV20}, not shown here, would approximate the early behavior well but has no mechanism to include NAC coupling.

\begin{figure}
    \centering
\includegraphics[width=0.5\textwidth]{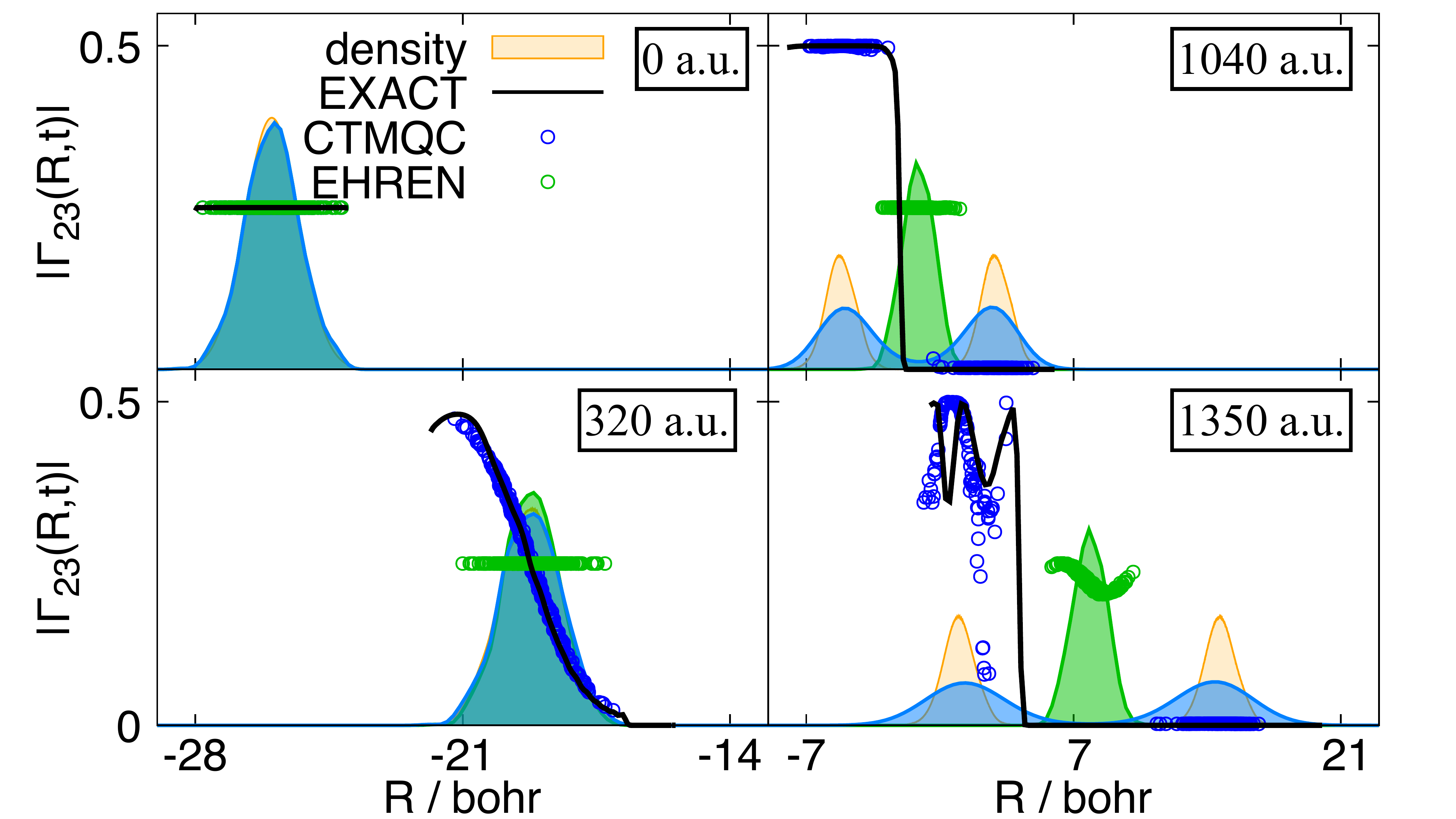}
    \caption{Snapshots of the exact density (orange shaded), density reconstructed from the distribution of CTMQC  (blue shaded) and {Ehrenfest} trajectories  (green shaded) and corresponding spatially-resolved coherences between the two parallel surfaces.}
    \label{fig:coh-el20sac_snap}
\end{figure}

Why the spatially-averaged electronic quantities are poor at later times in the traditional trajectory-based methods is revealed by inspecting the spatially-resolved quantities at earlier times, as plotted in Fig.~\ref{fig:coh-el20sac_snap}.
%{\color{purple} a movie is provided in the Supplementary Material; or shall we skip it?}. 
 Even before there is a clear splitting between the wavepackets on non-parallel surfaces (e.g. $t = 320$ a.u.), we see a distinct curvature in the exact spatially-resolved coherence (also in populations, not shown), captured well by CTMQC but completely missed in Ehrenfest which remains structureless until the avoided crossing region.
The projected quantum-momentum terms in the equation of motion  create this spatial structure; these terms are well-approximated by CTMQC but absent in the traditional methods, yet the latter correctly ensemble-average to zero before the SAC is reached. The incorrect spatial structure in Ehrenfest and SHEDC lead to incorrect ensemble-averaged populations and coherences at later times, since trajectories enter the SAC with wrong coefficients.

\begin{figure}
    \centering
\includegraphics[width=0.5\textwidth]{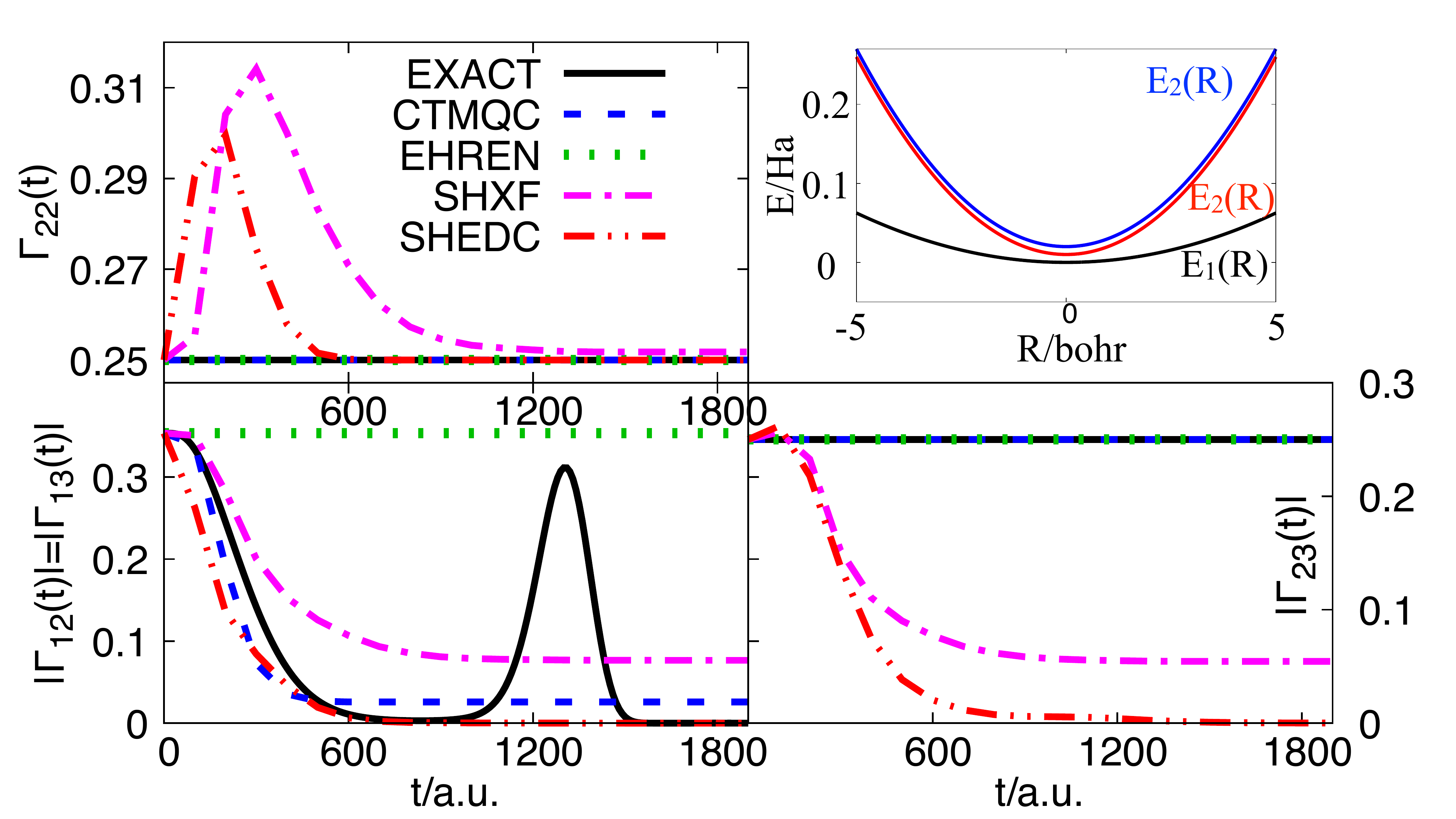}
    \caption{ As in Fig. 1, but for the 3HO model.}
  %  Electronic state populations of the first excited state (top left panel), adiabatic PES (top right panel), magnitude of electronic coherences between non parallel (bottom left panel) and parallel states (bottom right panel) as a function of time for 3HO model.}
    \label{fig:coh-tpargnu}
\end{figure}

The second model, denoted 3HO, consists of three uncoupled harmonic oscillators, two of which are parallel (inset of Fig.~\ref{fig:coh-tpargnu}); the Hamiltonian is given in the SM. The system is initialized in a $50:25:25$ coherent superposition of three gaussian nuclear wavepackets with zero momentum centered at $-4$ bohr.
%the same coherent superposition than for the EL20-SAC model (Eq.~\ref{eq:superpos}) with initial variance, position, momentum are $(\sigma,R_0,P_0)=(0.223\,\mathrm{bohr},-4\,\mathrm{bohr},0\,\hbar/\mathrm{bohr})$. 
While the wavepackets on the parallel surfaces maintain constant coherence, they show successive decoherences and recoherences with the wavepacket on the non-parallel surface. Figure~\ref{fig:coh-tpargnu} shows that, again, while Ehrenfest maintains coherence between all states, SHEDC shows decoherence for all. SHXF shows similar behavior to SHEDC, and both suffer from spurious population transfer~\cite{VAM22}. CTMQC on the other hand, correctly captures the first decoherence event while maintaining constant spatially-averaged coherences between the parallel surfaces, and correctly constant populations, due its redefinition of the quantum momentum preventing spurious net transfer. However, the recoherence is absent in all the methods, and would also be in the thawed Gaussian approach (not shown). This is because we enter the domain of the fourth observation made earlier and Eq.~(\ref{eq:recoherence}), where, after the first decoherence,  the $\mathfrak{S}_n$ are the only terms that can drive coherence change, and is missing in all approaches.  The importance of $\mathfrak{S}_n$ is evident from the snapshots of the exact $Q$, $\mathcal{Q}_1$ and $\mathfrak{S}_1$  shown in the SM (Fig. S.1), where $\mathfrak{S}$ displays
%  and, {\color{blue} $\sum_{\nu,n}\vert C_n\vert^2\mathfrak{S}_{\nu,n}\cdot\mathbf{f}_{\nu,n}=-(1/2)\sum_{\nu,n}\nabla_\nu\vert C_n\vert^2\cdot\mathbf{f}_{\nu,n}$ in 
%  the case where the population of a given electronic state for a given trajectory has collapsed to another state or set of states after the decoherence event, the EF-based methods lack a mechanism that induces a change in the magnitude of the electronic coherences (See Eq.~\ref{eq:dcoh/dt}) and Eq.~\ref{eq:dcohdt2}.). 
 large features that largely cancel those in $Q$, crunching it down to yield a relatively small $\mathcal{Q}_k$. 
 
 %{\color{blue} We note that while the thawed gaussian wavepacket approach~\cite{GVB20} would approximate the coherence evolution pre-SAC well, it has no mechanism to account for the NAC, and also has no unable to capture recoherence.}{\color{magenta}\it at least i think so}

%Figure~\ref{fig:crunch} shows time-snapshots of the nuclear quantum momentum $Q$, the projected quantity on state 1 $Q_1$ and the reduced $\mathfrak{S}_1$ during the first recoherence event. We observe that $\mathfrak{S}_1$ becomes relevant during this event and "crunches" the quantum momentum. A correction to the CTMQC algorithm could be proposed in which the term that induces recoherence, i.e $\sum_n\vert C_n\vert^2\mathfrak{S}_{\nu,n}\cdot\mathbf{f}_{\nu,n}$ is approximated via $\sum_n\vert C_n\vert^2\mathbf{Q}_{\nu}\cdot\mathbf{f}_{\nu,n}$.

In summary,  we identified the key role played by the projected nuclear quantum momentum, a signature of electron-nuclear correlation missed in earlier analyses, in determining the correct spatially-resolved electronic coherence, and the importance of spatial-resolution in influencing subsequent spatially-averaged coherences. The EF approach allows the definition of exact trajectory-based equations for the electronic coefficients, thanks to the notion of a single nuclear wavefunction with the correct density and current-density. Unlike traditional trajectory-based methods, the EF-based CTMQC is promising for simulating electronic coherences in molecules, because it approximates the projected nuclear quantum momentum, and accurately predicts coherences and populations in models in which traditional  trajectory-based methods fail, distinguishing coherence between parallel surfaces and decoherence between non-parallel.  
%The approximation however neglects a term that is needed to capture recoherence, and
Future work will explore a correction to CTMQC that reinstates the neglected $\mathfrak{S}_n$, which should then  capture recoherence, while simultaneously eliminating the need for redefining the quantum momentum.

\acknowledgments{We gratefully acknowledge financial aid from the National Science Foundation Award No. CHE-
2154829, and the Department of Energy, Office of Basic Energy Sciences,
Division of Chemical Sciences, Geosciences and Biosciences under
Award No. DE-SC0020044, the Computational Chemistry Center:
Chemistry in Solution and at Interfaces funded by the U.S. Department
of Energy, Office of Science Basic Energy Sciences, under
Award No. DE-SC0019394.}
\bibliography{ref_na}

%merlin.mbs apsrev4-1.bst 2010-07-25 4.21a (PWD, AO, DPC) hacked
%Control: key (0)
%Control: author (0) dotless jnrlst
%Control: editor formatted (1) identically to author
%Control: production of article title (0) allowed
%Control: page (1) range
%Control: year (0) verbatim
%Control: production of eprint (0) enabled
\begin{thebibliography}{52}%
\makeatletter
\providecommand \@ifxundefined [1]{%
 \@ifx{#1\undefined}
}%
\providecommand \@ifnum [1]{%
 \ifnum #1\expandafter \@firstoftwo
 \else \expandafter \@secondoftwo
 \fi
}%
\providecommand \@ifx [1]{%
 \ifx #1\expandafter \@firstoftwo
 \else \expandafter \@secondoftwo
 \fi
}%
\providecommand \natexlab [1]{#1}%
\providecommand \enquote  [1]{``#1''}%
\providecommand \bibnamefont  [1]{#1}%
\providecommand \bibfnamefont [1]{#1}%
\providecommand \citenamefont [1]{#1}%
\providecommand \href@noop [0]{\@secondoftwo}%
\providecommand \href [0]{\begingroup \@sanitize@url \@href}%
\providecommand \@href[1]{\@@startlink{#1}\@@href}%
\providecommand \@@href[1]{\endgroup#1\@@endlink}%
\providecommand \@sanitize@url [0]{\catcode `\\12\catcode `\$12\catcode
  `\&12\catcode `\#12\catcode `\^12\catcode `\_12\catcode `\%12\relax}%
\providecommand \@@startlink[1]{}%
\providecommand \@@endlink[0]{}%
\providecommand \url  [0]{\begingroup\@sanitize@url \@url }%
\providecommand \@url [1]{\endgroup\@href {#1}{\urlprefix }}%
\providecommand \urlprefix  [0]{URL }%
\providecommand \Eprint [0]{\href }%
\providecommand \doibase [0]{http://dx.doi.org/}%
\providecommand \selectlanguage [0]{\@gobble}%
\providecommand \bibinfo  [0]{\@secondoftwo}%
\providecommand \bibfield  [0]{\@secondoftwo}%
\providecommand \translation [1]{[#1]}%
\providecommand \BibitemOpen [0]{}%
\providecommand \bibitemStop [0]{}%
\providecommand \bibitemNoStop [0]{.\EOS\space}%
\providecommand \EOS [0]{\spacefactor3000\relax}%
\providecommand \BibitemShut  [1]{\csname bibitem#1\endcsname}%
\let\auto@bib@innerbib\@empty
%</preamble>
\bibitem [{\citenamefont {Kaufman}\ \emph {et~al.}(2023)\citenamefont
  {Kaufman}, \citenamefont {Marquetand}, \citenamefont {Rozgonyi},\ and\
  \citenamefont {Weinacht}}]{KMRW23b}%
  \BibitemOpen
  \bibfield  {author} {\bibinfo {author} {\bibfnamefont {Brian}\ \bibnamefont
  {Kaufman}}, \bibinfo {author} {\bibfnamefont {Philipp}\ \bibnamefont
  {Marquetand}}, \bibinfo {author} {\bibfnamefont {Tam\'as}\ \bibnamefont
  {Rozgonyi}}, \ and\ \bibinfo {author} {\bibfnamefont {Thomas}\ \bibnamefont
  {Weinacht}},\ }\bibfield  {title} {\enquote {\bibinfo {title} {Long-lived
  electronic coherences in molecules},}\ }\href {\doibase
  10.1103/PhysRevLett.131.263202} {\bibfield  {journal} {\bibinfo  {journal}
  {Phys. Rev. Lett.}\ }\textbf {\bibinfo {volume} {131}},\ \bibinfo {pages}
  {263202} (\bibinfo {year} {2023})}\BibitemShut {NoStop}%
\bibitem [{\citenamefont {Garg}\ \emph {et~al.}(2022)\citenamefont {Garg},
  \citenamefont {Martin-Jimenez}, \citenamefont {Pisarra}, \citenamefont {Luo},
  \citenamefont {Mart{\'i}n},\ and\ \citenamefont {Kern}}]{GMPLMK22}%
  \BibitemOpen
  \bibfield  {author} {\bibinfo {author} {\bibfnamefont {M.}~\bibnamefont
  {Garg}}, \bibinfo {author} {\bibfnamefont {A.}~\bibnamefont
  {Martin-Jimenez}}, \bibinfo {author} {\bibfnamefont {M.}~\bibnamefont
  {Pisarra}}, \bibinfo {author} {\bibfnamefont {Y.}~\bibnamefont {Luo}},
  \bibinfo {author} {\bibfnamefont {F.}~\bibnamefont {Mart{\'i}n}}, \ and\
  \bibinfo {author} {\bibfnamefont {K.}~\bibnamefont {Kern}},\ }\bibfield
  {title} {\enquote {\bibinfo {title} {Real-space subfemtosecond imaging of
  quantum electronic coherences in molecules},}\ }\href {\doibase
  10.1038/s41566-021-00929-1} {\bibfield  {journal} {\bibinfo  {journal}
  {Nature Photonics}\ }\textbf {\bibinfo {volume} {16}},\ \bibinfo {pages}
  {196--202} (\bibinfo {year} {2022})}\BibitemShut {NoStop}%
\bibitem [{\citenamefont {Mogol}\ \emph {et~al.}(2024)\citenamefont {Mogol},
  \citenamefont {Kaufman}, \citenamefont {Weinacht}, \citenamefont {Cheng},\
  and\ \citenamefont {Ben-Itzhak}}]{MKCBW23}%
  \BibitemOpen
  \bibfield  {author} {\bibinfo {author} {\bibfnamefont {G{\"o}nen~MC}\
  \bibnamefont {Mogol}}, \bibinfo {author} {\bibfnamefont {Brian}\ \bibnamefont
  {Kaufman}}, \bibinfo {author} {\bibfnamefont {Thomas}\ \bibnamefont
  {Weinacht}}, \bibinfo {author} {\bibfnamefont {Chuan}\ \bibnamefont {Cheng}},
  \ and\ \bibinfo {author} {\bibfnamefont {Itzik}\ \bibnamefont {Ben-Itzhak}},\
  }\bibfield  {title} {\enquote {\bibinfo {title} {Direct observation of
  entangled electronic-nuclear wave packets},}\ }\href {\doibase
  10.1103/PhysRevResearch.6.L022047} {\bibfield  {journal} {\bibinfo  {journal}
  {Phys. Rev. Res.}\ }\textbf {\bibinfo {volume} {6}},\ \bibinfo {pages}
  {L022047} (\bibinfo {year} {2024})}\BibitemShut {NoStop}%
\bibitem [{\citenamefont {Despr\'e}\ \emph {et~al.}(2018)\citenamefont
  {Despr\'e}, \citenamefont {Golubev},\ and\ \citenamefont {Kuleff}}]{DGK18}%
  \BibitemOpen
  \bibfield  {author} {\bibinfo {author} {\bibfnamefont {Victor}\ \bibnamefont
  {Despr\'e}}, \bibinfo {author} {\bibfnamefont {Nikolay~V.}\ \bibnamefont
  {Golubev}}, \ and\ \bibinfo {author} {\bibfnamefont {Alexander~I.}\
  \bibnamefont {Kuleff}},\ }\bibfield  {title} {\enquote {\bibinfo {title}
  {Charge migration in propiolic acid: A full quantum dynamical study},}\
  }\href {\doibase 10.1103/PhysRevLett.121.203002} {\bibfield  {journal}
  {\bibinfo  {journal} {Phys. Rev. Lett.}\ }\textbf {\bibinfo {volume} {121}},\
  \bibinfo {pages} {203002} (\bibinfo {year} {2018})}\BibitemShut {NoStop}%
\bibitem [{\citenamefont {Dey}\ \emph {et~al.}(2022)\citenamefont {Dey},
  \citenamefont {Kuleff},\ and\ \citenamefont {Worth}}]{DKW22}%
  \BibitemOpen
  \bibfield  {author} {\bibinfo {author} {\bibfnamefont {Diptesh}\ \bibnamefont
  {Dey}}, \bibinfo {author} {\bibfnamefont {Alexander~I.}\ \bibnamefont
  {Kuleff}}, \ and\ \bibinfo {author} {\bibfnamefont {Graham~A.}\ \bibnamefont
  {Worth}},\ }\bibfield  {title} {\enquote {\bibinfo {title} {Quantum
  interference paves the way for long-lived electronic coherences},}\ }\href
  {\doibase 10.1103/PhysRevLett.129.173203} {\bibfield  {journal} {\bibinfo
  {journal} {Phys. Rev. Lett.}\ }\textbf {\bibinfo {volume} {129}},\ \bibinfo
  {pages} {173203} (\bibinfo {year} {2022})}\BibitemShut {NoStop}%
\bibitem [{\citenamefont {Vacher}\ \emph {et~al.}(2017)\citenamefont {Vacher},
  \citenamefont {Bearpark}, \citenamefont {Robb},\ and\ \citenamefont
  {Malhado}}]{VBRM17}%
  \BibitemOpen
  \bibfield  {author} {\bibinfo {author} {\bibfnamefont {Morgane}\ \bibnamefont
  {Vacher}}, \bibinfo {author} {\bibfnamefont {Michael~J.}\ \bibnamefont
  {Bearpark}}, \bibinfo {author} {\bibfnamefont {Michael~A.}\ \bibnamefont
  {Robb}}, \ and\ \bibinfo {author} {\bibfnamefont {Jo\~ao~Pedro}\ \bibnamefont
  {Malhado}},\ }\bibfield  {title} {\enquote {\bibinfo {title} {Electron
  dynamics upon ionization of polyatomic molecules: Coupling to quantum nuclear
  motion and decoherence},}\ }\href {\doibase 10.1103/PhysRevLett.118.083001}
  {\bibfield  {journal} {\bibinfo  {journal} {Phys. Rev. Lett.}\ }\textbf
  {\bibinfo {volume} {118}},\ \bibinfo {pages} {083001} (\bibinfo {year}
  {2017})}\BibitemShut {NoStop}%
\bibitem [{\citenamefont {Duan}\ \emph {et~al.}(2017)\citenamefont {Duan},
  \citenamefont {Prokhorenko}, \citenamefont {Cogdell}, \citenamefont {Ashraf},
  \citenamefont {Stevens}, \citenamefont {Thorwart},\ and\ \citenamefont
  {Miller}}]{DPCASTM17}%
  \BibitemOpen
  \bibfield  {author} {\bibinfo {author} {\bibfnamefont {HG}~\bibnamefont
  {Duan}}, \bibinfo {author} {\bibfnamefont {VI}~\bibnamefont {Prokhorenko}},
  \bibinfo {author} {\bibfnamefont {RJ}~\bibnamefont {Cogdell}}, \bibinfo
  {author} {\bibfnamefont {K}~\bibnamefont {Ashraf}}, \bibinfo {author}
  {\bibfnamefont {AL}~\bibnamefont {Stevens}}, \bibinfo {author} {\bibfnamefont
  {M}~\bibnamefont {Thorwart}}, \ and\ \bibinfo {author} {\bibfnamefont {RJD}\
  \bibnamefont {Miller}},\ }\bibfield  {title} {\enquote {\bibinfo {title}
  {Nature does not rely on long-lived electronic quantum coherence for
  photosynthetic energy transfer},}\ }\href
  {https://doi.org/10.1073/pnas.1702261114} {\bibfield  {journal} {\bibinfo
  {journal} {Proc. Natl. Acad. Sci.}\ }\textbf {\bibinfo {volume} {114}}
  (\bibinfo {year} {2017})}\BibitemShut {NoStop}%
\bibitem [{\citenamefont {Maiuri}\ \emph {et~al.}(2018)\citenamefont {Maiuri},
  \citenamefont {Ostroumov}, \citenamefont {Saer}, \citenamefont
  {Blankenship},\ and\ \citenamefont {Scholes}}]{MOSBSG18}%
  \BibitemOpen
  \bibfield  {author} {\bibinfo {author} {\bibfnamefont {Margherita}\
  \bibnamefont {Maiuri}}, \bibinfo {author} {\bibfnamefont {Evgeny~E.}\
  \bibnamefont {Ostroumov}}, \bibinfo {author} {\bibfnamefont {Rafael~G.}\
  \bibnamefont {Saer}}, \bibinfo {author} {\bibfnamefont {Robert~E.}\
  \bibnamefont {Blankenship}}, \ and\ \bibinfo {author} {\bibfnamefont
  {Gregory~D.}\ \bibnamefont {Scholes}},\ }\bibfield  {title} {\enquote
  {\bibinfo {title} {Coherent wavepackets in the fenna--matthews--olson complex
  are robust to excitonic-structure perturbations caused by mutagenesis},}\
  }\href {https://doi.org/10.1038/nchem.2910} {\bibfield  {journal} {\bibinfo
  {journal} {Nat. Chem.}\ }\textbf {\bibinfo {volume} {10}},\ \bibinfo {pages}
  {177--183} (\bibinfo {year} {2018})}\BibitemShut {NoStop}%
\bibitem [{\citenamefont {Wasielewski}\ \emph {et~al.}(2020)\citenamefont
  {Wasielewski}, \citenamefont {Forbes}, \citenamefont {Frank}, \citenamefont
  {Kowalski}, \citenamefont {Scholes}, \citenamefont {Yuen-Zhou}, \citenamefont
  {Baldo}, \citenamefont {Freedman}, \citenamefont {Goldsmith}, \citenamefont
  {Goodson}, \citenamefont {Kirk}, \citenamefont {McCusker}, \citenamefont
  {Ogilvie}, \citenamefont {Shultz}, \citenamefont {Stoll},\ and\ \citenamefont
  {Whaley}}]{Wasielewski2020}%
  \BibitemOpen
  \bibfield  {author} {\bibinfo {author} {\bibfnamefont {Michael~R.}\
  \bibnamefont {Wasielewski}}, \bibinfo {author} {\bibfnamefont {Malcolm
  D.~E.}\ \bibnamefont {Forbes}}, \bibinfo {author} {\bibfnamefont {Natia~L.}\
  \bibnamefont {Frank}}, \bibinfo {author} {\bibfnamefont {Karol}\ \bibnamefont
  {Kowalski}}, \bibinfo {author} {\bibfnamefont {Gregory~D.}\ \bibnamefont
  {Scholes}}, \bibinfo {author} {\bibfnamefont {Joel}\ \bibnamefont
  {Yuen-Zhou}}, \bibinfo {author} {\bibfnamefont {Marc~A.}\ \bibnamefont
  {Baldo}}, \bibinfo {author} {\bibfnamefont {Danna~E.}\ \bibnamefont
  {Freedman}}, \bibinfo {author} {\bibfnamefont {Randall~H.}\ \bibnamefont
  {Goldsmith}}, \bibinfo {author} {\bibfnamefont {Theodore}\ \bibnamefont
  {Goodson}}, \bibinfo {author} {\bibfnamefont {Martin~L.}\ \bibnamefont
  {Kirk}}, \bibinfo {author} {\bibfnamefont {James~K.}\ \bibnamefont
  {McCusker}}, \bibinfo {author} {\bibfnamefont {Jennifer~P.}\ \bibnamefont
  {Ogilvie}}, \bibinfo {author} {\bibfnamefont {David~A.}\ \bibnamefont
  {Shultz}}, \bibinfo {author} {\bibfnamefont {Stefan}\ \bibnamefont {Stoll}},
  \ and\ \bibinfo {author} {\bibfnamefont {K.~Birgitta}\ \bibnamefont
  {Whaley}},\ }\bibfield  {title} {\enquote {\bibinfo {title} {Exploiting
  chemistry and molecular systems for quantum information science},}\ }\href
  {\doibase 10.1038/s41570-020-0200-5} {\bibfield  {journal} {\bibinfo
  {journal} {Nature Reviews Chemistry}\ }\textbf {\bibinfo {volume} {4}},\
  \bibinfo {pages} {490--504} (\bibinfo {year} {2020})}\BibitemShut {NoStop}%
\bibitem [{\citenamefont {Villaseco~Arribas}\ \emph {et~al.}(2024)\citenamefont
  {Villaseco~Arribas}, \citenamefont {Maitra},\ and\ \citenamefont
  {Agostini}}]{VMA24}%
  \BibitemOpen
  \bibfield  {author} {\bibinfo {author} {\bibfnamefont {Evaristo}\
  \bibnamefont {Villaseco~Arribas}}, \bibinfo {author} {\bibfnamefont
  {Neepa~T.}\ \bibnamefont {Maitra}}, \ and\ \bibinfo {author} {\bibfnamefont
  {Federica}\ \bibnamefont {Agostini}},\ }\bibfield  {title} {\enquote
  {\bibinfo {title} {Nonadiabatic dynamics with classical trajectories: The
  problem of an initial coherent superposition of electronic states},}\ }\href
  {\doibase 10.1063/5.0186984} {\bibfield  {journal} {\bibinfo  {journal} {J.
  Chem. Phys.}\ }\textbf {\bibinfo {volume} {160}},\ \bibinfo {pages} {054102}
  (\bibinfo {year} {2024})},\ \Eprint
  {http://arxiv.org/abs/https://pubs.aip.org/aip/jcp/article-pdf/doi/10.1063/5.0186984/19333721/054102\_1\_5.0186984.pdf}
  {https://pubs.aip.org/aip/jcp/article-pdf/doi/10.1063/5.0186984/19333721/054102\_1\_5.0186984.pdf}
  \BibitemShut {NoStop}%
\bibitem [{\citenamefont {Li}\ \emph {et~al.}(2020)\citenamefont {Li},
  \citenamefont {Gyawali}, \citenamefont {Ischenko}, \citenamefont {Hayes},\
  and\ \citenamefont {Miller}}]{LGIHM20}%
  \BibitemOpen
  \bibfield  {author} {\bibinfo {author} {\bibfnamefont {Zheng}\ \bibnamefont
  {Li}}, \bibinfo {author} {\bibfnamefont {Sandeep}\ \bibnamefont {Gyawali}},
  \bibinfo {author} {\bibfnamefont {Anatoly~A.}\ \bibnamefont {Ischenko}},
  \bibinfo {author} {\bibfnamefont {Stuart}\ \bibnamefont {Hayes}}, \ and\
  \bibinfo {author} {\bibfnamefont {R.~J.~Dwayne}\ \bibnamefont {Miller}},\
  }\bibfield  {title} {\enquote {\bibinfo {title} {Mapping atomic motions with
  electrons: Toward the quantum limit to imaging chemistry},}\ }\href {\doibase
  10.1021/acsphotonics.9b01008} {\bibfield  {journal} {\bibinfo  {journal} {ACS
  Photonics}\ }\textbf {\bibinfo {volume} {7}},\ \bibinfo {pages} {296--320}
  (\bibinfo {year} {2020})},\ \Eprint
  {http://arxiv.org/abs/https://doi.org/10.1021/acsphotonics.9b01008}
  {https://doi.org/10.1021/acsphotonics.9b01008} \BibitemShut {NoStop}%
\bibitem [{\citenamefont {Cavaletto}\ \emph {et~al.}(2021)\citenamefont
  {Cavaletto}, \citenamefont {Keefer}, \citenamefont {Rouxel}, \citenamefont
  {Aleotti}, \citenamefont {Segatta}, \citenamefont {Garavelli},\ and\
  \citenamefont {Mukamel}}]{CKRASGM21}%
  \BibitemOpen
  \bibfield  {author} {\bibinfo {author} {\bibfnamefont {SM}~\bibnamefont
  {Cavaletto}}, \bibinfo {author} {\bibfnamefont {D}~\bibnamefont {Keefer}},
  \bibinfo {author} {\bibfnamefont {JR}~\bibnamefont {Rouxel}}, \bibinfo
  {author} {\bibfnamefont {F}~\bibnamefont {Aleotti}}, \bibinfo {author}
  {\bibfnamefont {F}~\bibnamefont {Segatta}}, \bibinfo {author} {\bibfnamefont
  {M}~\bibnamefont {Garavelli}}, \ and\ \bibinfo {author} {\bibfnamefont
  {S.}~\bibnamefont {Mukamel}},\ }\bibfield  {title} {\enquote {\bibinfo
  {title} {Unveiling the spatial distribution of molecular coherences at
  conical intersections by covariance x-ray diffraction signals},}\ }\href
  {https://doi.org/10.1073/pnas.2105046118} {\bibfield  {journal} {\bibinfo
  {journal} {Proc. Natl. Acad. Sci.}\ }\textbf {\bibinfo {volume} {118}}
  (\bibinfo {year} {2021})}\BibitemShut {NoStop}%
\bibitem [{\citenamefont {Fiete}\ and\ \citenamefont {Heller}(2003)}]{FH03}%
  \BibitemOpen
  \bibfield  {author} {\bibinfo {author} {\bibfnamefont {Gregory~A.}\
  \bibnamefont {Fiete}}\ and\ \bibinfo {author} {\bibfnamefont {Eric~J.}\
  \bibnamefont {Heller}},\ }\bibfield  {title} {\enquote {\bibinfo {title}
  {Semiclassical theory of coherence and decoherence},}\ }\href {\doibase
  10.1103/PhysRevA.68.022112} {\bibfield  {journal} {\bibinfo  {journal} {Phys.
  Rev. A}\ }\textbf {\bibinfo {volume} {68}},\ \bibinfo {pages} {022112}
  (\bibinfo {year} {2003})}\BibitemShut {NoStop}%
\bibitem [{\citenamefont {Golubev}\ \emph {et~al.}(2020)\citenamefont
  {Golubev}, \citenamefont {Begu\ifmmode \check{s}\else
  \v{s}\fi{}i\ifmmode~\acute{c}\else \'{c}\fi{}},\ and\ \citenamefont
  {Van\'{\i}\ifmmode~\check{c}\else \v{c}\fi{}ek}}]{GBV20}%
  \BibitemOpen
  \bibfield  {author} {\bibinfo {author} {\bibfnamefont {Nikolay~V.}\
  \bibnamefont {Golubev}}, \bibinfo {author} {\bibfnamefont {Tomislav}\
  \bibnamefont {Begu\ifmmode \check{s}\else \v{s}\fi{}i\ifmmode~\acute{c}\else
  \'{c}\fi{}}}, \ and\ \bibinfo {author} {\bibfnamefont {Ji\ifmmode
  \check{r}\else~\v{r}\fi{}\'{\i}}\ \bibnamefont
  {Van\'{\i}\ifmmode~\check{c}\else \v{c}\fi{}ek}},\ }\bibfield  {title}
  {\enquote {\bibinfo {title} {On-the-fly ab initio semiclassical evaluation of
  electronic coherences in polyatomic molecules reveals a simple mechanism of
  decoherence},}\ }\href {\doibase 10.1103/PhysRevLett.125.083001} {\bibfield
  {journal} {\bibinfo  {journal} {Phys. Rev. Lett.}\ }\textbf {\bibinfo
  {volume} {125}},\ \bibinfo {pages} {083001} (\bibinfo {year}
  {2020})}\BibitemShut {NoStop}%
\bibitem [{\citenamefont {Tully}(1990)}]{T90}%
  \BibitemOpen
  \bibfield  {author} {\bibinfo {author} {\bibfnamefont {J.~C.}\ \bibnamefont
  {Tully}},\ }\bibfield  {title} {\enquote {\bibinfo {title} {{Molecular
  dynamics with electronic transitions}},}\ }\href@noop {} {\bibfield
  {journal} {\bibinfo  {journal} {J. Chem. Phys.}\ }\textbf {\bibinfo {volume}
  {93}},\ \bibinfo {pages} {1061--1071} (\bibinfo {year} {1990})}\BibitemShut
  {NoStop}%
\bibitem [{\citenamefont {Tully}(1998)}]{T98}%
  \BibitemOpen
  \bibfield  {author} {\bibinfo {author} {\bibfnamefont {J.~C.}\ \bibnamefont
  {Tully}},\ }\bibfield  {title} {\enquote {\bibinfo {title} {Mixed
  quantum-classical dynamics},}\ }\href@noop {} {\bibfield  {journal} {\bibinfo
   {journal} {Faraday Discuss.}\ }\textbf {\bibinfo {volume} {110}},\ \bibinfo
  {pages} {407--419} (\bibinfo {year} {1998})}\BibitemShut {NoStop}%
\bibitem [{\citenamefont {Wang}\ \emph {et~al.}(2016)\citenamefont {Wang},
  \citenamefont {Akimov},\ and\ \citenamefont {Prezhdo}}]{WAP16}%
  \BibitemOpen
  \bibfield  {author} {\bibinfo {author} {\bibfnamefont {Linjun}\ \bibnamefont
  {Wang}}, \bibinfo {author} {\bibfnamefont {Alexey}\ \bibnamefont {Akimov}}, \
  and\ \bibinfo {author} {\bibfnamefont {Oleg~V.}\ \bibnamefont {Prezhdo}},\
  }\bibfield  {title} {\enquote {\bibinfo {title} {{Recent Progress in Surface
  Hopping: 2011-2015}},}\ }\href@noop {} {\bibfield  {journal} {\bibinfo
  {journal} {J. Phys. Chem. Lett.}\ }\textbf {\bibinfo {volume} {7}},\ \bibinfo
  {pages} {2100--2112} (\bibinfo {year} {2016})}\BibitemShut {NoStop}%
\bibitem [{\citenamefont {Crespo-Otero}\ and\ \citenamefont
  {Barbatti}(2018)}]{CB18}%
  \BibitemOpen
  \bibfield  {author} {\bibinfo {author} {\bibfnamefont {Rachel}\ \bibnamefont
  {Crespo-Otero}}\ and\ \bibinfo {author} {\bibfnamefont {Mario}\ \bibnamefont
  {Barbatti}},\ }\bibfield  {title} {\enquote {\bibinfo {title} {{Recent
  Advances and Perspectives on Nonadiabatic Mixed Quantum--Classical
  Dynamics}},}\ }\href@noop {} {\bibfield  {journal} {\bibinfo  {journal}
  {Chem. Rev.}\ }\textbf {\bibinfo {volume} {118}},\ \bibinfo {pages}
  {7026--7068} (\bibinfo {year} {2018})}\BibitemShut {NoStop}%
\bibitem [{\citenamefont {Subotnik}\ \emph {et~al.}(2016)\citenamefont
  {Subotnik}, \citenamefont {Jain}, \citenamefont {Landry}, \citenamefont
  {Petit}, \citenamefont {Ouyang},\ and\ \citenamefont {Bellonzi}}]{SJLP16}%
  \BibitemOpen
  \bibfield  {author} {\bibinfo {author} {\bibfnamefont {Joseph~E.}\
  \bibnamefont {Subotnik}}, \bibinfo {author} {\bibfnamefont {Amber}\
  \bibnamefont {Jain}}, \bibinfo {author} {\bibfnamefont {Brian}\ \bibnamefont
  {Landry}}, \bibinfo {author} {\bibfnamefont {Andrew}\ \bibnamefont {Petit}},
  \bibinfo {author} {\bibfnamefont {Wenjun}\ \bibnamefont {Ouyang}}, \ and\
  \bibinfo {author} {\bibfnamefont {Nicole}\ \bibnamefont {Bellonzi}},\
  }\bibfield  {title} {\enquote {\bibinfo {title} {{Understanding the Surface
  Hopping View of Electronic Transitions and Decoherence}},}\ }\href@noop {}
  {\bibfield  {journal} {\bibinfo  {journal} {Ann. Rev. Phys. Chem.}\ }\textbf
  {\bibinfo {volume} {67}},\ \bibinfo {pages} {387--417} (\bibinfo {year}
  {2016})}\BibitemShut {NoStop}%
\bibitem [{\citenamefont {Esch}\ and\ \citenamefont {Levine}(2020)}]{EL20}%
  \BibitemOpen
  \bibfield  {author} {\bibinfo {author} {\bibfnamefont {Michael~P.}\
  \bibnamefont {Esch}}\ and\ \bibinfo {author} {\bibfnamefont {Benjamin~G.}\
  \bibnamefont {Levine}},\ }\bibfield  {title} {\enquote {\bibinfo {title}
  {{State-pairwise decoherence times for nonadiabatic dynamics on more than two
  electronic states}},}\ }\href {https://doi.org/10.1063/5.0010081} {\bibfield
  {journal} {\bibinfo  {journal} {J. Chem. Phys.}\ }\textbf {\bibinfo {volume}
  {152}},\ \bibinfo {pages} {234105} (\bibinfo {year} {2020})}\BibitemShut
  {NoStop}%
\bibitem [{\citenamefont {Min}\ \emph {et~al.}(2015)\citenamefont {Min},
  \citenamefont {Agostini},\ and\ \citenamefont {Gross}}]{MAG15}%
  \BibitemOpen
  \bibfield  {author} {\bibinfo {author} {\bibfnamefont {Seung~Kyu}\
  \bibnamefont {Min}}, \bibinfo {author} {\bibfnamefont {Federica}\
  \bibnamefont {Agostini}}, \ and\ \bibinfo {author} {\bibfnamefont {E.~K.~U.}\
  \bibnamefont {Gross}},\ }\bibfield  {title} {\enquote {\bibinfo {title}
  {{Coupled-Trajectory Quantum-Classical Approach to Electronic Decoherence in
  Nonadiabatic Processes}},}\ }\href@noop {} {\bibfield  {journal} {\bibinfo
  {journal} {Phys. Rev. Lett.}\ }\textbf {\bibinfo {volume} {115}},\ \bibinfo
  {pages} {073001} (\bibinfo {year} {2015})}\BibitemShut {NoStop}%
\bibitem [{\citenamefont {Agostini}\ \emph {et~al.}(2016)\citenamefont
  {Agostini}, \citenamefont {Min}, \citenamefont {Abedi},\ and\ \citenamefont
  {Gross}}]{AMAG16}%
  \BibitemOpen
  \bibfield  {author} {\bibinfo {author} {\bibfnamefont {Federica}\
  \bibnamefont {Agostini}}, \bibinfo {author} {\bibfnamefont {Seung~Kyu}\
  \bibnamefont {Min}}, \bibinfo {author} {\bibfnamefont {Ali}\ \bibnamefont
  {Abedi}}, \ and\ \bibinfo {author} {\bibfnamefont {E.~K.~U.}\ \bibnamefont
  {Gross}},\ }\bibfield  {title} {\enquote {\bibinfo {title}
  {{Quantum-Classical Nonadiabatic Dynamics: Coupled- vs Independent-Trajectory
  Methods}},}\ }\href@noop {} {\bibfield  {journal} {\bibinfo  {journal} {J.
  Chem. Theory Comput.}\ }\textbf {\bibinfo {volume} {12}},\ \bibinfo {pages}
  {2127--2143} (\bibinfo {year} {2016})}\BibitemShut {NoStop}%
\bibitem [{\citenamefont {Ha}\ \emph {et~al.}(2018)\citenamefont {Ha},
  \citenamefont {Lee},\ and\ \citenamefont {Min}}]{HLM18}%
  \BibitemOpen
  \bibfield  {author} {\bibinfo {author} {\bibfnamefont {Jong-Kwon}\
  \bibnamefont {Ha}}, \bibinfo {author} {\bibfnamefont {In~Seong}\ \bibnamefont
  {Lee}}, \ and\ \bibinfo {author} {\bibfnamefont {Seung~Kyu}\ \bibnamefont
  {Min}},\ }\bibfield  {title} {\enquote {\bibinfo {title} {{Surface Hopping
  Dynamics beyond Nonadiabatic Couplings for Quantum Coherence}},}\ }\href@noop
  {} {\bibfield  {journal} {\bibinfo  {journal} {J. Phys. Chem. Lett.}\
  }\textbf {\bibinfo {volume} {9}},\ \bibinfo {pages} {1097--1104} (\bibinfo
  {year} {2018})}\BibitemShut {NoStop}%
\bibitem [{\citenamefont {Villaseco~Arribas}\ and\ \citenamefont
  {Maitra}(2023)}]{VM23}%
  \BibitemOpen
  \bibfield  {author} {\bibinfo {author} {\bibfnamefont {Evaristo}\
  \bibnamefont {Villaseco~Arribas}}\ and\ \bibinfo {author} {\bibfnamefont
  {Neepa~T.}\ \bibnamefont {Maitra}},\ }\bibfield  {title} {\enquote {\bibinfo
  {title} {{Energy-conserving coupled trajectory mixed quantum-classical
  dynamics}},}\ }\href@noop {} {\bibfield  {journal} {\bibinfo  {journal} {J.
  Chem. Phys.}\ }\textbf {\bibinfo {volume} {158}},\ \bibinfo {pages} {161105}
  (\bibinfo {year} {2023})}\BibitemShut {NoStop}%
\bibitem [{\citenamefont {Dupuy}\ \emph {et~al.}(2024)\citenamefont {Dupuy},
  \citenamefont {Rikus},\ and\ \citenamefont {Maitra}}]{DRM24}%
  \BibitemOpen
  \bibfield  {author} {\bibinfo {author} {\bibfnamefont {Lucien}\ \bibnamefont
  {Dupuy}}, \bibinfo {author} {\bibfnamefont {Anton}\ \bibnamefont {Rikus}}, \
  and\ \bibinfo {author} {\bibfnamefont {Neepa~T.}\ \bibnamefont {Maitra}},\
  }\bibfield  {title} {\enquote {\bibinfo {title} {Exact-factorization-based
  surface hopping without velocity adjustment},}\ }\href
  {https://doi.org/10.1021/acs.jpclett.4c00115} {\bibfield  {journal} {\bibinfo
   {journal} {The Journal of Physical Chemistry Letters}\ }\textbf {\bibinfo
  {volume} {15}},\ \bibinfo {pages} {2643--2649} (\bibinfo {year}
  {2024})}\BibitemShut {NoStop}%
\bibitem [{\citenamefont {Li}\ \emph {et~al.}(2022)\citenamefont {Li},
  \citenamefont {Requist},\ and\ \citenamefont {Gross}}]{LRG22}%
  \BibitemOpen
  \bibfield  {author} {\bibinfo {author} {\bibfnamefont {Chen}\ \bibnamefont
  {Li}}, \bibinfo {author} {\bibfnamefont {Ryan}\ \bibnamefont {Requist}}, \
  and\ \bibinfo {author} {\bibfnamefont {E.~K.~U.}\ \bibnamefont {Gross}},\
  }\bibfield  {title} {\enquote {\bibinfo {title} {Energy, momentum, and
  angular momentum transfer between electrons and nuclei},}\ }\href {\doibase
  10.1103/PhysRevLett.128.113001} {\bibfield  {journal} {\bibinfo  {journal}
  {Phys. Rev. Lett.}\ }\textbf {\bibinfo {volume} {128}},\ \bibinfo {pages}
  {113001} (\bibinfo {year} {2022})}\BibitemShut {NoStop}%
\bibitem [{\citenamefont {Agostini}\ \emph {et~al.}(2014)\citenamefont
  {Agostini}, \citenamefont {Abedi},\ and\ \citenamefont {Gross}}]{AAG14}%
  \BibitemOpen
  \bibfield  {author} {\bibinfo {author} {\bibfnamefont {F.}~\bibnamefont
  {Agostini}}, \bibinfo {author} {\bibfnamefont {A.}~\bibnamefont {Abedi}}, \
  and\ \bibinfo {author} {\bibfnamefont {E.~K.~U.}\ \bibnamefont {Gross}},\
  }\bibfield  {title} {\enquote {\bibinfo {title} {{Classical nuclear motion
  coupled to electronic non-adiabatic transitions}},}\ }\href@noop {}
  {\bibfield  {journal} {\bibinfo  {journal} {J. Chem. Phys.}\ }\textbf
  {\bibinfo {volume} {141}},\ \bibinfo {pages} {214101} (\bibinfo {year}
  {2014})}\BibitemShut {NoStop}%
\bibitem [{\citenamefont {Abedi}\ \emph {et~al.}(2014)\citenamefont {Abedi},
  \citenamefont {Agostini},\ and\ \citenamefont {Gross}}]{AAG14b}%
  \BibitemOpen
  \bibfield  {author} {\bibinfo {author} {\bibfnamefont {A.}~\bibnamefont
  {Abedi}}, \bibinfo {author} {\bibfnamefont {F.}~\bibnamefont {Agostini}}, \
  and\ \bibinfo {author} {\bibfnamefont {E.~K.~U.}\ \bibnamefont {Gross}},\
  }\bibfield  {title} {\enquote {\bibinfo {title} {{Mixed quantum-classical
  dynamics from the exact decomposition of electron-nuclear motion}},}\
  }\href@noop {} {\bibfield  {journal} {\bibinfo  {journal} {Europhys. Lett.}\
  }\textbf {\bibinfo {volume} {106}},\ \bibinfo {pages} {33001} (\bibinfo
  {year} {2014})}\BibitemShut {NoStop}%
\bibitem [{\citenamefont {Hunter}(1975{\natexlab{a}})}]{Hunter75}%
  \BibitemOpen
  \bibfield  {author} {\bibinfo {author} {\bibfnamefont {Geoffrey}\
  \bibnamefont {Hunter}},\ }\bibfield  {title} {\enquote {\bibinfo {title}
  {{Conditional probability amplitudes in wave mechanics}},}\ }\href@noop {}
  {\bibfield  {journal} {\bibinfo  {journal} {Int. J. Quantum Chem.}\ }\textbf
  {\bibinfo {volume} {9}},\ \bibinfo {pages} {237} (\bibinfo {year}
  {1975}{\natexlab{a}})}\BibitemShut {NoStop}%
\bibitem [{\citenamefont {Hunter}(1975{\natexlab{b}})}]{Hunter_IJQC1975_2}%
  \BibitemOpen
  \bibfield  {author} {\bibinfo {author} {\bibfnamefont {Geoffrey}\
  \bibnamefont {Hunter}},\ }\bibfield  {title} {\enquote {\bibinfo {title}
  {{Ionization potentials and conditional amplitudes}},}\ }\href@noop {}
  {\bibfield  {journal} {\bibinfo  {journal} {Int. J. Quantum Chem.}\ }\textbf
  {\bibinfo {volume} {9}},\ \bibinfo {pages} {311} (\bibinfo {year}
  {1975}{\natexlab{b}})}\BibitemShut {NoStop}%
\bibitem [{\citenamefont {Hunter}(1980)}]{Hunter_IJQC1980}%
  \BibitemOpen
  \bibfield  {author} {\bibinfo {author} {\bibfnamefont {Geoffrey}\
  \bibnamefont {Hunter}},\ }\bibfield  {title} {\enquote {\bibinfo {title}
  {{Nodeless wave function quantum theory}},}\ }\href@noop {} {\bibfield
  {journal} {\bibinfo  {journal} {Int. J. Quantum Chem.}\ }\textbf {\bibinfo
  {volume} {9}},\ \bibinfo {pages} {133} (\bibinfo {year} {1980})}\BibitemShut
  {NoStop}%
\bibitem [{\citenamefont {Hunter}(1981)}]{H81}%
  \BibitemOpen
  \bibfield  {author} {\bibinfo {author} {\bibfnamefont {Geoffrey}\
  \bibnamefont {Hunter}},\ }\bibfield  {title} {\enquote {\bibinfo {title}
  {{Nodeless wave functions and spiky potentials}},}\ }\href@noop {} {\bibfield
   {journal} {\bibinfo  {journal} {Int. J. Quantum Chem.}\ }\textbf {\bibinfo
  {volume} {19}},\ \bibinfo {pages} {755} (\bibinfo {year} {1981})}\BibitemShut
  {NoStop}%
\bibitem [{\citenamefont {Hunter}\ and\ \citenamefont
  {Tai}(1982)}]{Hunter_IJQC1982}%
  \BibitemOpen
  \bibfield  {author} {\bibinfo {author} {\bibfnamefont {Geoffrey}\
  \bibnamefont {Hunter}}\ and\ \bibinfo {author} {\bibfnamefont {Chin~Chui}\
  \bibnamefont {Tai}},\ }\bibfield  {title} {\enquote {\bibinfo {title}
  {{Variational marginal amplitudes}},}\ }\href@noop {} {\bibfield  {journal}
  {\bibinfo  {journal} {Int. J. Quantum Chem.}\ }\textbf {\bibinfo {volume}
  {21}},\ \bibinfo {pages} {1041} (\bibinfo {year} {1982})}\BibitemShut
  {NoStop}%
\bibitem [{\citenamefont {Gidopoulos}\ and\ \citenamefont
  {Gross}(2014)}]{GG14}%
  \BibitemOpen
  \bibfield  {author} {\bibinfo {author} {\bibfnamefont {Nikitas~I.}\
  \bibnamefont {Gidopoulos}}\ and\ \bibinfo {author} {\bibfnamefont {E.~K.~U.}\
  \bibnamefont {Gross}},\ }\bibfield  {title} {\enquote {\bibinfo {title}
  {{Electronic non-adiabatic states: towards a density functional theory beyond
  the Born{\textendash}Oppenheimer approximation}},}\ }\href@noop {} {\bibfield
   {journal} {\bibinfo  {journal} {Philosophical Transactions of the Royal
  Society of London A: Mathematical, Physical and Engineering Sciences}\
  }\textbf {\bibinfo {volume} {372}} (\bibinfo {year} {2014})}\BibitemShut
  {NoStop}%
\bibitem [{\citenamefont {Abedi}\ \emph {et~al.}(2010)\citenamefont {Abedi},
  \citenamefont {Maitra},\ and\ \citenamefont {Gross}}]{AMG10}%
  \BibitemOpen
  \bibfield  {author} {\bibinfo {author} {\bibfnamefont {Ali}\ \bibnamefont
  {Abedi}}, \bibinfo {author} {\bibfnamefont {Neepa~T.}\ \bibnamefont
  {Maitra}}, \ and\ \bibinfo {author} {\bibfnamefont {E.~K.~U.}\ \bibnamefont
  {Gross}},\ }\bibfield  {title} {\enquote {\bibinfo {title} {{Exact
  Factorization of the Time-Dependent Electron-Nuclear Wave Function}},}\
  }\href@noop {} {\bibfield  {journal} {\bibinfo  {journal} {Phys. Rev. Lett.}\
  }\textbf {\bibinfo {volume} {105}},\ \bibinfo {pages} {123002} (\bibinfo
  {year} {2010})}\BibitemShut {NoStop}%
\bibitem [{\citenamefont {Abedi}\ \emph {et~al.}(2012)\citenamefont {Abedi},
  \citenamefont {Maitra},\ and\ \citenamefont {Gross}}]{AMG12}%
  \BibitemOpen
  \bibfield  {author} {\bibinfo {author} {\bibfnamefont {Ali}\ \bibnamefont
  {Abedi}}, \bibinfo {author} {\bibfnamefont {Neepa~T.}\ \bibnamefont
  {Maitra}}, \ and\ \bibinfo {author} {\bibfnamefont {E.~K.~U.}\ \bibnamefont
  {Gross}},\ }\bibfield  {title} {\enquote {\bibinfo {title} {{Correlated
  electron-nuclear dynamics: Exact factorization of the molecular
  wavefunction}},}\ }\href@noop {} {\bibfield  {journal} {\bibinfo  {journal}
  {J. Chem. Phys.}\ }\textbf {\bibinfo {volume} {137}},\ \bibinfo {pages}
  {22A530} (\bibinfo {year} {2012})}\BibitemShut {NoStop}%
\bibitem [{\citenamefont {Agostini}\ and\ \citenamefont {Gross}(2021)}]{AG21}%
  \BibitemOpen
  \bibfield  {author} {\bibinfo {author} {\bibfnamefont {Federica}\
  \bibnamefont {Agostini}}\ and\ \bibinfo {author} {\bibfnamefont {E.~K.~U.}\
  \bibnamefont {Gross}},\ }\bibfield  {title} {\enquote {\bibinfo {title}
  {{Ultrafast dynamics with the exact factorization}},}\ }\href@noop {}
  {\bibfield  {journal} {\bibinfo  {journal} {Eur. Phys. J. B}\ }\textbf
  {\bibinfo {volume} {94}},\ \bibinfo {pages} {179} (\bibinfo {year}
  {2021})}\BibitemShut {NoStop}%
\bibitem [{\citenamefont {Villaseco~Arribas}\ \emph {et~al.}(2022)\citenamefont
  {Villaseco~Arribas}, \citenamefont {Agostini},\ and\ \citenamefont
  {Maitra}}]{VAM22}%
  \BibitemOpen
  \bibfield  {author} {\bibinfo {author} {\bibfnamefont {Evaristo}\
  \bibnamefont {Villaseco~Arribas}}, \bibinfo {author} {\bibfnamefont
  {Federica}\ \bibnamefont {Agostini}}, \ and\ \bibinfo {author} {\bibfnamefont
  {Neepa~T.}\ \bibnamefont {Maitra}},\ }\bibfield  {title} {\enquote {\bibinfo
  {title} {{Exact Factorization Adventures: A Promising Approach for Non-Bound
  States}},}\ }\href@noop {} {\bibfield  {journal} {\bibinfo  {journal}
  {Molecules}\ }\textbf {\bibinfo {volume} {27}},\ \bibinfo {pages} {13}
  (\bibinfo {year} {2022})}\BibitemShut {NoStop}%
\bibitem [{Note1()}]{Note1}%
  \BibitemOpen
  \bibinfo {note} {{See Supplemental Material which includes Refs.~\cite
  {Hunter75,Hunter_IJQC1975_2,Hunter_IJQC1980,H81,Hunter_IJQC1982,GG14,AMG10,AMG12,AG21,VAM22,AAG14,AAG14b,MAG15,ATC18,LRG22,WRB04,BM98,BQM00,S09,MSFS17,MGMS14,GCTMQC,PyUNIxMD,VMA24}
  }}\BibitemShut {NoStop}%
\bibitem [{\citenamefont {Min}\ \emph {et~al.}(2017)\citenamefont {Min},
  \citenamefont {Agostini}, \citenamefont {Tavernelli},\ and\ \citenamefont
  {Gross}}]{MATG17}%
  \BibitemOpen
  \bibfield  {author} {\bibinfo {author} {\bibfnamefont {Seung~Kyu}\
  \bibnamefont {Min}}, \bibinfo {author} {\bibfnamefont {Federica}\
  \bibnamefont {Agostini}}, \bibinfo {author} {\bibfnamefont {Ivano}\
  \bibnamefont {Tavernelli}}, \ and\ \bibinfo {author} {\bibfnamefont
  {E.~K.~U.}\ \bibnamefont {Gross}},\ }\bibfield  {title} {\enquote {\bibinfo
  {title} {{Ab Initio Nonadiabatic Dynamics with Coupled Trajectories: A
  Rigorous Approach to Quantum (De)Coherence}},}\ }\href@noop {} {\bibfield
  {journal} {\bibinfo  {journal} {J. Phys. Chem. Lett.}\ }\textbf {\bibinfo
  {volume} {8}},\ \bibinfo {pages} {3048--3055} (\bibinfo {year}
  {2017})}\BibitemShut {NoStop}%
\bibitem [{Note2()}]{Note2}%
  \BibitemOpen
  \bibinfo {note} {{Note that here we have shown $\DOTSI \intop \ilimits@ \vert
  \Gamma _{ij}(R,t) \vert dR$ as the magnitude of the coherence, rather than
  $\vert \DOTSI \intop \ilimits@ \Gamma _{ij}(R,t) dR \vert $, i.e. the measure
  we are showing does not include dephasing.}}\BibitemShut {Stop}%
\bibitem [{\citenamefont {Granucci}\ and\ \citenamefont
  {Persico}(2007)}]{GP07}%
  \BibitemOpen
  \bibfield  {author} {\bibinfo {author} {\bibfnamefont {G.}~\bibnamefont
  {Granucci}}\ and\ \bibinfo {author} {\bibfnamefont {M.}~\bibnamefont
  {Persico}},\ }\bibfield  {title} {\enquote {\bibinfo {title} {{Critical
  appraisal of the fewest switches algorithm for surface hopping}},}\
  }\href@noop {} {\bibfield  {journal} {\bibinfo  {journal} {J. Chem. Phys.}\
  }\textbf {\bibinfo {volume} {126}},\ \bibinfo {pages} {134114} (\bibinfo
  {year} {2007})}\BibitemShut {NoStop}%
\bibitem [{\citenamefont {Granucci}\ \emph {et~al.}(2010)\citenamefont
  {Granucci}, \citenamefont {Persico},\ and\ \citenamefont
  {Zoccante}}]{GPZ2010}%
  \BibitemOpen
  \bibfield  {author} {\bibinfo {author} {\bibfnamefont {Giovanni}\
  \bibnamefont {Granucci}}, \bibinfo {author} {\bibfnamefont {Maurizio}\
  \bibnamefont {Persico}}, \ and\ \bibinfo {author} {\bibfnamefont {Alberto}\
  \bibnamefont {Zoccante}},\ }\bibfield  {title} {\enquote {\bibinfo {title}
  {{Including quantum decoherence in surface hopping}},}\ }\href@noop {}
  {\bibfield  {journal} {\bibinfo  {journal} {J. Chem. Phys.}\ }\textbf
  {\bibinfo {volume} {133}},\ \bibinfo {pages} {134111} (\bibinfo {year}
  {2010})}\BibitemShut {NoStop}%
\bibitem [{\citenamefont {Lee}\ \emph {et~al.}(2021)\citenamefont {Lee},
  \citenamefont {Ha}, \citenamefont {Han}, \citenamefont {Kim}, \citenamefont
  {Moon},\ and\ \citenamefont {Min}}]{PyUNIxMD}%
  \BibitemOpen
  \bibfield  {author} {\bibinfo {author} {\bibfnamefont {In~Seong}\
  \bibnamefont {Lee}}, \bibinfo {author} {\bibfnamefont {Jong-Kwon}\
  \bibnamefont {Ha}}, \bibinfo {author} {\bibfnamefont {Daeho}\ \bibnamefont
  {Han}}, \bibinfo {author} {\bibfnamefont {Tae~In}\ \bibnamefont {Kim}},
  \bibinfo {author} {\bibfnamefont {Sung~Wook}\ \bibnamefont {Moon}}, \ and\
  \bibinfo {author} {\bibfnamefont {Seung~Kyu}\ \bibnamefont {Min}},\
  }\bibfield  {title} {\enquote {\bibinfo {title} {{PyUNIxMD: A Python-based
  excited state molecular dynamics package}},}\ }\href@noop {} {\bibfield
  {journal} {\bibinfo  {journal} {J. Comput. Chem.}\ }\textbf {\bibinfo
  {volume} {42}},\ \bibinfo {pages} {1755--1766} (\bibinfo {year}
  {2021})}\BibitemShut {NoStop}%
\bibitem [{\citenamefont {Agostini}\ \emph {et~al.}(2018)\citenamefont
  {Agostini}, \citenamefont {Tavernelli},\ and\ \citenamefont
  {Ciccotti}}]{ATC18}%
  \BibitemOpen
  \bibfield  {author} {\bibinfo {author} {\bibfnamefont {F.}~\bibnamefont
  {Agostini}}, \bibinfo {author} {\bibfnamefont {I.}~\bibnamefont
  {Tavernelli}}, \ and\ \bibinfo {author} {\bibfnamefont {G.}~\bibnamefont
  {Ciccotti}},\ }\bibfield  {title} {\enquote {\bibinfo {title} {{Nuclear
  Quantum Effects in Electronic (Non)Adiabatic Dynamics}},}\ }\href@noop {}
  {\bibfield  {journal} {\bibinfo  {journal} {Euro. Phys. J. B}\ }\textbf
  {\bibinfo {volume} {91}},\ \bibinfo {pages} {139} (\bibinfo {year}
  {2018})}\BibitemShut {NoStop}%
\bibitem [{\citenamefont {Worth}\ \emph {et~al.}(2004)\citenamefont {Worth},
  \citenamefont {Robb},\ and\ \citenamefont {Burghardt}}]{WRB04}%
  \BibitemOpen
  \bibfield  {author} {\bibinfo {author} {\bibfnamefont {G.~A.}\ \bibnamefont
  {Worth}}, \bibinfo {author} {\bibfnamefont {M.~A.}\ \bibnamefont {Robb}}, \
  and\ \bibinfo {author} {\bibfnamefont {I.}~\bibnamefont {Burghardt}},\
  }\bibfield  {title} {\enquote {\bibinfo {title} {A novel algorithm for
  non-adiabatic direct dynamics using variational gaussian wavepackets},}\
  }\href {\doibase 10.1039/B314253A} {\bibfield  {journal} {\bibinfo  {journal}
  {Faraday Discuss.}\ }\textbf {\bibinfo {volume} {127}},\ \bibinfo {pages}
  {307--323} (\bibinfo {year} {2004})}\BibitemShut {NoStop}%
\bibitem [{\citenamefont {Ben-Nun}\ and\ \citenamefont
  {Martinez}(1998)}]{BM98}%
  \BibitemOpen
  \bibfield  {author} {\bibinfo {author} {\bibfnamefont {M.}~\bibnamefont
  {Ben-Nun}}\ and\ \bibinfo {author} {\bibfnamefont {Todd~J.}\ \bibnamefont
  {Martinez}},\ }\bibfield  {title} {\enquote {\bibinfo {title} {{Nonadiabatic
  molecular dynamics: Validation of the multiple spawning method for a
  multidimensional problem}},}\ }\href@noop {} {\bibfield  {journal} {\bibinfo
  {journal} {J. Chem. Phys.}\ }\textbf {\bibinfo {volume} {108}},\ \bibinfo
  {pages} {7244--7257} (\bibinfo {year} {1998})}\BibitemShut {NoStop}%
\bibitem [{\citenamefont {Ben-Nun}\ \emph {et~al.}(2000)\citenamefont
  {Ben-Nun}, \citenamefont {Quenneville},\ and\ \citenamefont
  {Mart{\'i}nez}}]{BQM00}%
  \BibitemOpen
  \bibfield  {author} {\bibinfo {author} {\bibfnamefont {M.}~\bibnamefont
  {Ben-Nun}}, \bibinfo {author} {\bibfnamefont {Jason}\ \bibnamefont
  {Quenneville}}, \ and\ \bibinfo {author} {\bibfnamefont {Todd~J.}\
  \bibnamefont {Mart{\'i}nez}},\ }\bibfield  {title} {\enquote {\bibinfo
  {title} {{Ab Initio Multiple Spawning: Photochemistry from First Principles
  Quantum Molecular Dynamics}},}\ }\href@noop {} {\bibfield  {journal}
  {\bibinfo  {journal} {J. Phys. Chem. A}\ }\textbf {\bibinfo {volume} {104}},\
  \bibinfo {pages} {5161--5175} (\bibinfo {year} {2000})}\BibitemShut {NoStop}%
\bibitem [{\citenamefont {Shalashilin}(2009)}]{S09}%
  \BibitemOpen
  \bibfield  {author} {\bibinfo {author} {\bibfnamefont {Dmitrii~V.}\
  \bibnamefont {Shalashilin}},\ }\bibfield  {title} {\enquote {\bibinfo {title}
  {Quantum mechanics with the basis set guided by ehrenfest trajectories:
  Theory and application to spin-boson model},}\ }\href {\doibase
  10.1063/1.3153302} {\bibfield  {journal} {\bibinfo  {journal} {J. Chem.
  Phys.}\ }\textbf {\bibinfo {volume} {130}},\ \bibinfo {pages} {244101}
  (\bibinfo {year} {2009})}\BibitemShut {NoStop}%
\bibitem [{\citenamefont {Makhov}\ \emph {et~al.}(2017)\citenamefont {Makhov},
  \citenamefont {Symonds}, \citenamefont {Fernandez-Alberti},\ and\
  \citenamefont {Shalashilin}}]{MSFS17}%
  \BibitemOpen
  \bibfield  {author} {\bibinfo {author} {\bibfnamefont {Dmitry~V.}\
  \bibnamefont {Makhov}}, \bibinfo {author} {\bibfnamefont {Christopher}\
  \bibnamefont {Symonds}}, \bibinfo {author} {\bibfnamefont {Sebastian}\
  \bibnamefont {Fernandez-Alberti}}, \ and\ \bibinfo {author} {\bibfnamefont
  {Dmitrii~V.}\ \bibnamefont {Shalashilin}},\ }\bibfield  {title} {\enquote
  {\bibinfo {title} {Ab initio quantum direct dynamics simulations of ultrafast
  photochemistry with multiconfigurational ehrenfest approach},}\ }\href@noop
  {} {\bibfield  {journal} {\bibinfo  {journal} {Chemical Physics}\ }\textbf
  {\bibinfo {volume} {493}},\ \bibinfo {pages} {200--218} (\bibinfo {year}
  {2017})}\BibitemShut {NoStop}%
\bibitem [{\citenamefont {Makhov}\ \emph {et~al.}(2014)\citenamefont {Makhov},
  \citenamefont {Glover}, \citenamefont {Martinez},\ and\ \citenamefont
  {Shalashilin}}]{MGMS14}%
  \BibitemOpen
  \bibfield  {author} {\bibinfo {author} {\bibfnamefont {Dmitry~V.}\
  \bibnamefont {Makhov}}, \bibinfo {author} {\bibfnamefont {William~J.}\
  \bibnamefont {Glover}}, \bibinfo {author} {\bibfnamefont {Todd~J.}\
  \bibnamefont {Martinez}}, \ and\ \bibinfo {author} {\bibfnamefont
  {Dmitrii~V.}\ \bibnamefont {Shalashilin}},\ }\bibfield  {title} {\enquote
  {\bibinfo {title} {{Ab initio multiple cloning algorithm for quantum
  nonadiabatic molecular dynamics}},}\ }\href {\doibase 10.1063/1.4891530}
  {\bibfield  {journal} {\bibinfo  {journal} {J. Chem. Phys.}\ }\textbf
  {\bibinfo {volume} {141}},\ \bibinfo {pages} {054110} (\bibinfo {year}
  {2014})}\BibitemShut {NoStop}%
\bibitem [{\citenamefont {Agostini}\ \emph {et~al.}(Last accessed Oct
  2023)\citenamefont {Agostini}, \citenamefont {Marsili}, \citenamefont
  {Talotta}, \citenamefont {Villaseco~Arribas}, \citenamefont {Ibele},\ and\
  \citenamefont {Sangiogo~Gil}}]{GCTMQC}%
  \BibitemOpen
  \bibfield  {author} {\bibinfo {author} {\bibfnamefont {Federica}\
  \bibnamefont {Agostini}}, \bibinfo {author} {\bibfnamefont {Emmanuele}\
  \bibnamefont {Marsili}}, \bibinfo {author} {\bibfnamefont {Francesco}\
  \bibnamefont {Talotta}}, \bibinfo {author} {\bibfnamefont {Evaristo}\
  \bibnamefont {Villaseco~Arribas}}, \bibinfo {author} {\bibfnamefont
  {Lea~Maria}\ \bibnamefont {Ibele}}, \ and\ \bibinfo {author} {\bibfnamefont
  {Eduarda.}\ \bibnamefont {Sangiogo~Gil}},\ }\href@noop {} {\enquote {\bibinfo
  {title} {{{G-CTMQC}}},}\ } (\bibinfo {year} {Last accessed Oct
  2023})\BibitemShut {NoStop}%
\end{thebibliography}%

\onecolumngrid
\newpage
\renewcommand{\theequation}{S.\arabic{equation}}
\renewcommand{\thefigure}{S.1}

\section{Supplementary Material}
\section{SM.1 Exact factorization of the molecular wavefunction}
The full time-dependent molecular wavefunction can be exactly factorized as a single correlated product~\cite{Hunter75,Hunter_IJQC1975_2,Hunter_IJQC1980,H81,Hunter_IJQC1982,GG14,AMG10,AMG12,AG21,VAM22}, $\Psi(\dulr,\dulR,t) = \chi(\dulR,t)\Phi_\dulR(\dulr,t)$ where the electronic wavefunction satisfies the partial normalization condition $\int \vert\Phi_\dulR(\dulr,t)\vert^2 d\dulr =1\,\forall\,\dulR,t$. The marginal factor $\chi(\dulR,t)=e^{iS(\dulR,t)}\vert\chi(\dulR,t)\vert$ is interpreted as a nuclear wavefunction because its modulus-square gives the exact nuclear density and the gradient of its phase the exact nuclear current-density (and nuclear velocity field) of the molecular wavefunction:
\bea
\vert \chi(\dulR,t) \vert^2 &=& \int \vert \Psi(\dulr,\dulR,t) \vert^2 d\dulr  \;\;\;\\
\mathbf{V}(\dulR,t)=\frac{\nabla_\nu S(\dulR,t)}{M_\nu}&=&\frac{{\bf J}_\nu(\dulR,t)}{\vert \chi(\dulR,t)\vert^2 } -\frac{{\bf A}_\nu(\dulR,t)}{M_\nu}\label{eq:grad_phase}
\eea
with $M_\nu{\bf J}_\nu(\dulR,t)=Im\langle\Psi( \dulR,t)\vert\nabla_\nu \Psi(\dulR,t)\rangle_\dulr$.  The time-evolution for the electronic and nuclear subsystems satisfy
\begin{eqnarray}
i\partial_t\chi(\dulR,t)&=&\hat{H}_N(\dulR,t)\chi(\dulR,t) \\
i\partial_t\Phi_\dulR(\dulr,t)&=&\left[\hat{H}_{el}(\dulr,\dulR)-\epsilon(\dulR,t)\right]\Phi_\dulR(\dulr,t)
\end{eqnarray}
The nuclear equation is of Schr\"odinger form with Hamiltonian 
\begin{equation}
    \hat{H}_N(\dulR,t)=\sum_\nu\frac{(-i\nabla_\nu+\mathbf{A}_\nu(\dulR,t))^2}{M_\nu}+\epsilon(\dulR,t)\,.
    \end{equation}
The scalar $\epsilon(\dulR,t)=\langle \Phi_\dulR(t)\vert\hat{H}_{el}-i\partial_t\vert\Phi_\dulR(t)\rangle_\dulr$ and vector potentials $\mathbf{A}_\nu(\dulR,t)=-i\langle \Phi_\dulR(t)\vert\nabla_\nu\Phi_\dulR(t)\rangle_\dulr$ incorporate the effect of the electronic wavefunction in the nuclear subsystem. On the other hand, the electronic 
equation is non-linear with electronic Hamiltonian defined as
\begin{equation}
 \hat{H}_{el}(\dulr,\dulR)=\hat{H}_{BO}(\dulr;\dulR)+\hat{U}_{eN}[\Phi_\dulR,\chi] 
\end{equation}
and electron-nuclear coupling operator defined as
 \begin{equation}
 \hat{U}_{eN}=\sum_\nu\frac{1}{M_\nu}\left[\frac{\left(-i\nabla_\nu-\mathbf{A}_\nu(\dulR,t)\right)^2}{2}+\left(\frac{-i\nabla_\nu\chi(\dulR,t)}{\chi(\dulR,t)}+\mathbf{A}_\nu(\dulR,t)\right)\left(-i\nabla_\nu-\mathbf{A}_\nu(\dulR,t)\right)\right]    
 \end{equation}
 which incorporates the effect of the nuclear wavefunction in the electronic subsystem. 
 
 Note that the factorized form of $\Psi(\dulr,\dulR,t)$ is exact and unique up to a $\dulR$ and $t$ dependent phase, i.e $\widetilde{\chi}(\dulR,t)\rightarrow e^{i\theta(\dulR,t)}\chi(\dulR,t)$, $\widetilde{\Phi}_\dulR(\dulr,t)\rightarrow e^{-i\theta(\dulR,t)}\Phi_\dulR(\dulr,t)$, $\widetilde{\Psi}(\dulr,\dulR,t)\rightarrow\Psi(\dulr,\dulR,t)$, with $\epsilon(\dulR,t)$ and $\mathbf{A}_\nu(\dulR,t)$ transforming according to standard electrodynamic potentials $\widetilde{\mathbf{A}}_\nu(\dulR,t)\rightarrow \mathbf{A}_\nu(\dulR,t)+\nabla_\nu\theta(\dulR,t)$ and $\widetilde{\epsilon}(\dulR,t)=\epsilon(\dulR,t)+\partial_t \theta(\dulR,t)$.

 Although the single-product form of the exact factorization (EF) approach resembles the form of the molecular wavefunction in the Born-Oppenheimer approximation,  a significant distinction is that in the former, the equations for the electronic and nuclear wavefunctions must be solved self-consistently, while in the latter, the electronic equation can be solved first for each $\dulR$ and then the nuclear equation solved.

While both the Born-Huang expansion and the exact factorization approach are formally exact, 
the advantage of using the exact factorization approach for these problems is that it enables the definition of a unique unambiguous force~\cite{AAG14,AAG14b,MAG15,ATC18,LRG22} that drives nuclear trajectories in mixed quantum-classical (MQC) schemes. This force incorporates coupling to electrons (and any external fields) in a rigorous way. While the Born-Huang expansion also gives a formally exact representation of the  molecular wavefunction, it is more challenging to derive a consistent set of equations in MQC schemes because nuclear wavepackets on different electronic surfaces experience different forces, which gives ambiguity to the force on a classical trajectory at a given nuclear position.
For example, Ehrenfest simply takes an average of the adiabatic forces, weighted by the electronic coefficients, but this mean-field force cannot yield wavepacket splitting, becoming increasingly unphysical away from regions of strong electron-nuclear coupling (avoided or conical intersections). To avoid this, the surface hopping scheme instead uses an adiabatic force  at all times, but to account for coupling must stochastically hop between active states. While there are more sophisticated trajectory-based schemes stemming from the Born-Huang expansion (e.g. variational multiconfiguration Gaussian~\cite{WRB04}, ab initio multiple spawning~\cite{BM98,BQM00},  multi-configurational Ehrenfest~\cite{S09,MSFS17}, ab initio multiple cloning~\cite{MGMS14}), the simplicity and physicality of a single unique nuclear force defined over the ensemble of classical trajectories, with each trajectory associated with quantum electronic coefficients,  gives EF-based MQC a particular advantage, conceptually as well as numerically.

%\section{SI.2 Classical-like nuclear force}
%{\color{purple} do we actually want/need to give this? because we woul dhave to then explain what the gradients mean for along the Lagrangian trajectory...}
%The classical trajectory is evolved via Newton’s equation
%\begin{equation}
% M_\nu\ddot\bR_\nu^{(I)} =\mathbf{E}_\nu[\Phi_\dulR]+\dot{\mathbf{R}}_\nu\times\mathbf{B}_{\nu\nu}[\Phi_\dulR] +\sum_{\mu\neq\nu}\mathbf{F}^\prime_{\nu\mu}[\Phi_\dulR]   
%\end{equation}
%where $\mathbf{E}_\nu=-\nabla_\nu\epsilon-\partial_t\mathbf{A}_\nu$ is an effective electric field and $\mathbf{B}_{\nu\mu}=\nabla_\nu\times\bold{A}_\mu$ is a generalized magnetic field. Hence, the force driving the nuclei is a generalized Lorentz force with an inter-nuclear force term, namely $\mathbf{F}^\prime_{\nu\mu}=-\dot{\mathbf{R}}_\mu\times\mathbf{B}_{\nu\mu}+(\dot{\mathbf{R}}_\mu\cdot\nabla_\mu)\bold{A}_\nu-(\dot{\mathbf{R}}_\mu\cdot\nabla_\nu)\bold{A}_\mu$, which  arises from the coupling with the electronic subsystem~\cite{AAG14,AAG14b,LRG22}.

\section{SM.2 Exact equations for the evolution of electronic coherences}
The electronic coherence defined in Eq. (1) of the main text, is based on the projected nuclear wavefunctions $\chi_j(\dulR,t)$ in the Born-Huang basis, $\Psi(\dulr,\dulR,t) \stackrel{\rm BO}{=} \sum_j\chi_j(\dulR,t)\Phi_{\dulR,j}(\dulr)$. 
The full, exact, equation for the time evolution of the magnitude of the electronic coherences of Eq.(1), including non-adiabatic couplings, reads:
\begin{eqnarray}
i \partial_t \Gamma_{jk}(\dulR,t) &=& \Delta E_{kj}(\dulR)\Gamma_{jk}(\dulR,t)+\sum_\nu \frac{1}{2M_\nu}\bigg(\chi_k(\dulR,t)\nabla_\nu^2\chi_j^* (\dulR,t) -\chi_j^*(\dulR,t)\nabla_\nu^2\chi_k(\dulR,t)\nonumber \\
&&-\sum_{n\neq k}\left[D_{kn,\nu}(\dulR)\Gamma_{jn}(\dulR,t)+2\chi_j^*(\dulR,t)\bold{d}_{kn,\nu}(\dulR)\cdot\nabla_\nu\chi_n(\dulR,t)\right]\nonumber \\
&&+\sum_{n\neq j}\left[D_{nj,\nu}(\dulR)\Gamma_{nk}(\dulR,t)+2\chi_k\bold{d}_{jn,\nu}(\dulR)\cdot\nabla_\nu\chi_n^*(\dulR,t)\right]\Bigg)
\end{eqnarray}
where we have assumed real BO states,  no geometric phase effects, $\Delta E_{kj} = E_k - E_j$, and
where $\bold{d}_{nk,\nu}(\dulR)$ and $D_{nk,\nu}(\dulR)$ are the non-adiabatic coupling vectors $\langle\phi_{\dulR, n}\vert\nabla_\nu\phi_{\dulR.k}\rangle_\dulr$ and scalar couplings $\langle\phi_{\dulR, n}\vert\nabla_\nu^2\phi_{\dulR, k}\rangle_\dulr$ respectively. Writing the wavepackets in polar form, i.e $\chi_n(\dulR,t) = \vert \chi_n(\dulR,t)\vert e^{iS_n(\dulR,t)}$, we have 
\begin{eqnarray}
\nabla_\nu\chi_k(\dulR,t)
&=&\left(i\nabla_\nu S_k(\dulR,t)-\pq_{\nu,k}(\dulR,t)\right)\chi_k(\dulR,t) \label{eq:nablachi}\\
\nabla_\nu^2\chi_k(\dulR,t)
&=&\left(\mathfrak{Q}_{\nu,k}(\dulR,t)-2i\pq_{\nu,k}(\dulR,t)\cdot\nabla_\nu S_k(\dulR,t)+i\nabla_\nu^2 S_k(\dulR,t) -(\nabla_\nu S_k(\dulR,t))^2\right)\chi_k(\dulR,t)\label{eq:nabla2chi}
\end{eqnarray}
where the k-th state projected quantum momentum and quantum potential read
\begin{equation}
 \pq_{\nu,k}(\dulR,t)=- \frac{\nabla_\nu\vert\chi_k(\dulR,t)\vert}{\vert\chi_k(\dulR,t)\vert}\,\,\,\,\,\,\,\,\,\,\,\,\,\,\,\,\,{\rm and } \,\,\,\,\,\,\,\,\,\,\,\,\,\,\,\,\,  \mathfrak{Q}_{\nu,k}(\dulR,t)=\frac{\nabla_\nu^2\vert\chi_k(\dulR,t)\vert}{\vert\chi_k(\dulR,t)\vert}\,.
\end{equation}
We now write these quantities in terms of the exact factorization  concepts, where we note that 
\ben
\Psi(\dulr,\dulR,t) \stackrel{\rm BO}{=} \sum_k\chi_j(\dulR,t)\Phi_{\dulR,k}(\dulr) \stackrel{\rm EF}{=} \chi(\dulR,t)\Phi^{\rm BO}_\dulR(\dulr,t) = \chi(\dulR,t)\sum_k C_k(\dulR,t)\Phi^{\rm BO}_{\dulR,k}(\dulr) 
\een
where in the last equality the time-dependent conditional electronic wavefunction is represented by its coefficients $\{C_k\}$ in the expansion, which also represent the projected nuclear wavefunction via:
\ben
\Phi_\dulR(\dulr,t) = \sum_k C_k(\dulR,t)\Phi^{\rm BO}_{\dulR,k}(\dulr)\,\,\, \,\,\, {\rm and}\,\,\,\,\,\, \chi_k(\dulR,t)=\chi(\dulR,t)C_k(\dulR,t)\,.
\label{eq:cond_exp}
\een
Writing the nuclear wavefunction $\chi$ and the coefficients $C_k$ also in polar form, 
$\chi(\dulR,t) = \vert \chi(\dulR,t) \vert e^{iS(\dulR,t)}$ and $C_k(\dulR,t) = \vert C_k(\dulR,t)\vert e^{i\gamma_k}$, Eq.~\ref{eq:cond_exp} means we can relate
\begin{equation}
S_k(\dulR,t) = S(\dulR,t) + \gamma_k(\dulR,t) \;\;\;\;\; {\rm and}\;\;\;\;\; \vert \chi_k(\dulR,t)\vert = \vert\chi(\dulR,t)\vert\vert C_k(\dulR,t)\vert
%\chi_k(\dulR,t)=\chi(\dulR,t)C_k(\dulR,t)\,\,\,\,\rightarrow e^{iS_k(\dulR,t)}\vert\chi_k(\dulR,t)\vert=e^{i\left[S(\dulR,t)+\gamma_k(\dulR,t)\right]}\vert\chi(\dulR,t)\vert\vert C_k(\dulR,t)\vert \,.
\end{equation}

We define the $k$-th state reduced quantum momentum as
\begin{equation}
 \mathfrak{S}_{\nu,k}(\dulR,t)=- \frac{\nabla_\nu\vert C_k(\dulR,t)\vert}{\vert C_k(\dulR,t)\vert}
\end{equation}
which allows us to write, equivalently to Eq~\ref{eq:nablachi},
\begin{equation}
\nabla_\nu C_k(\dulR,t)
=\left(i\nabla_\nu \gamma_k(\dulR,t)-\mathfrak{S}_{\nu,k}(\dulR,t)\right) C_k(\dulR,t) \label{eq:nablac}
\end{equation}
Note that the reduced and projected quantum momenta, $\mathfrak{S}_{\nu,k}(\dulR,t)$ and $\pq_{\nu,k}(\dulR,t)$, are related through the nuclear quantum momentum $\mathbf{Q}_\nu(\dulR,t)$
\begin{equation}
\mathbf{Q}_\nu(\dulR,t)=-\frac{\nabla_\nu\vert\chi(\dulR,t)\vert}{\vert\chi(\dulR,t)\vert}=\mathfrak{S}_{\nu,k}(\dulR,t)+\pq_{\nu,k}(\dulR,t)\,\,\,\,\,\,\,\,\forall\,\,\nu,k  
\label{eq:crunch}
\end{equation}
In terms of these quantities, we write, omitting the dependencies to avoid notational clutter,
\begin{eqnarray}
&&i \partial_t \Gamma_{jk}(\dulR,t)= \Delta E_{kj}\Gamma_{jk}+\sum_\nu \frac{\Gamma_{jk}}{2M_\nu}\bigg(\Delta\mathfrak{Q}_{\nu,jk}+2i\pq_{\nu,j}\nabla_\nu S_j-i\nabla_\nu^2 S_j -(\nabla_\nu S_j)^2+2i\pq_{\nu,k}\nabla_\nu S_k-i\nabla_\nu^2 S_k+(\nabla_\nu S_k)^2\Bigg) \nonumber \\
&&-\sum_{\nu,n\neq k}\frac{\Gamma_{jn}}{2M_\nu}\left[D_{kn,\nu}+2i\bold{d}_{kn,\nu}\cdot\nabla_\nu S_n-2\bold{d}_{kn,\nu}\cdot\pq_{\nu,n})\right]+\sum_{\nu,n\neq j}\frac{\Gamma_{nk}}{2M_\nu}\left[D_{nj,\nu}-2i\bold{d}_{jn,\nu}\cdot\nabla_\nu S_n-2\bold{d}_{jn,\nu}\cdot\pq_{\nu,n}\right]\Bigg)\nonumber \\
\label{eq:dGamRtdt2}
\end{eqnarray}

%\begin{eqnarray}
%&&i \partial_t \Gamma_{jk}(\dulR,t)= \Delta E_{kj}\Gamma_{jk}+\sum_\nu \frac{\Gamma_{jk}}{2M_\nu}\bigg(\mathfrak{Q}_{\nu,j}+2i\pq_{\nu,j}\nabla_\nu S_j-i\nabla_\nu^2 S_j -(\nabla_\nu S_j)^2-\mathfrak{Q}_{\nu,k}+2i\pq_{\nu,k}\nabla_\nu S_k-i\nabla_\nu^2 S_k+(\nabla_\nu S_k)^2\Bigg) \nonumber \\
%&&-\sum_{\nu,n\neq k}\frac{\Gamma_{jn}}{2M_\nu}\left[D_{kn,\nu}+2i\bold{d}_{kn,\nu}\cdot\nabla_\nu S_n-2\bold{d}_{kn,\nu}\cdot\pq_{\nu,n})\right]+\sum_{\nu,n\neq j}\frac{\Gamma_{nk}}{2M_\nu}\left[D_{nj,\nu}-2i\bold{d}_{jn,\nu}\cdot\nabla_\nu S_n-2\bold{d}_{jn,\nu}\cdot\pq_{\nu,n}\right]\Bigg)\nonumber \\
%\label{eq:dGamRtdt2}
%\end{eqnarray}
where we define $\Delta g_{jk} = g_j - g_k$ for any quantity. The advantage of writing  in terms of exact factorization quantities  means we can relate terms to the nuclear density and current-density (with an eventual view towards establishing a mixed quantum-classical method):  from Eq.~\ref{eq:grad_phase}, the inline equation for $\bf A_\nu$ below Eq. (S.5), and the expansion in Eq.~\ref{eq:cond_exp}, we have 
\begin{equation}
\nabla_\nu S=M_\nu\mathbf{V}_\nu=\frac{M_\nu\bold{J}_\nu}{\vert\chi\vert^2}-\bold{A}_\nu=\frac{M_\nu\bold{J}_\nu}{\vert\chi\vert^2}-\sum_l\vert C_l\vert^2\nabla_\nu\gamma_{l}- {\rm Im} \sum_{kl} C_l^*C_k {\bf d}_{\nu,lk}
\end{equation}
and
\begin{eqnarray}
\nabla_\nu ^2S&=&M_\nu\bold{J}_\nu\cdot\nabla_\nu\frac{1}{\vert\chi\vert^2}+\frac{M_\nu\nabla_\nu\cdot\bold{J}_\nu}{\vert\chi\vert^2}-\nabla_\nu\cdot\bold{A}_\nu \nonumber \\
&=&\frac{M_\nu}{\vert\chi\vert^2}\left(\nabla_\nu\cdot\bold{J}_\nu+2\left(\pq_{\nu,k}-\mathfrak{S}_{\nu,k}\right)\cdot\bold{J}_\nu \right)-\sum_l\vert C_l\vert^2\left(\nabla_\nu^2\gamma_l-2\mathfrak{S}_{\nu,l}\nabla_\nu\gamma_l\right) - {\rm Im} \nabla_\nu \cdot \sum_{kl} C_l^*C_k {\bf d}_{\nu,lk}
\end{eqnarray}
where we wrote $\bQ_\nu = \pq_{\nu,k}-\mathfrak{S}_{\nu,k}$ (Eq.~(\ref{eq:crunch})). In regions where the NACs are zero, this gives
%{\color{blue}
\begin{eqnarray}
\partial_t \Gamma_{jk}(\dulR,t) &=& -i\Delta E_{kj}(\dulR)\Gamma_{jk}(\dulR,t)+\sum_\nu \frac{\Gamma_{jk}}{2M_\nu}\bigg[-i\Delta\mathfrak{Q}_{\nu,jk}+2\left(\pq_{\nu,j}+\pq_{\nu,k}\right)\cdot\left(\cancel{\frac{M_\nu\bold{J}_\nu}{\vert\chi(\dulR,t)\vert^2}}-\sum_l\vert C_l\vert^2\nabla_\nu\gamma_{l}   \right) \nonumber \\
&&+2\left(\pq_{\nu,j}\cdot\nabla_\nu\gamma_j+\pq_{\nu,k}\cdot\nabla_\nu\gamma_k\right)-\frac{2M_\nu}{\vert\chi(\dulR,t)\vert^2}\left(\nabla_\nu\cdot\bold{J}_\nu+\left(\cancel{\pq_{\nu,k}+\pq_{\nu,j}}-\mathfrak{S}_{\nu,k}-\mathfrak{S}_{\nu,j}\right)\cdot\bold{J}_\nu \right) \nonumber \\
&&+2\sum_l\vert C_l\vert^2\left(\nabla_\nu^2\gamma_l-2\mathfrak{S}_{\nu,l}\nabla_\nu\gamma_l\right)-\nabla_\nu^2\left(\gamma_{k}+\gamma_{j}\right)-i(\nabla_\nu\gamma_k)^2+i(\nabla_\nu\gamma_j)^2 \nonumber \\
&&-2i\Delta\nabla_\nu\gamma_{kj}\cdot\left(\frac{M_\nu\bold{J}_\nu}{\vert\chi(\dulR,t)\vert^2}-\sum_l\vert C_l\vert^2\nabla_\nu\gamma_{l}   \right)\Bigg]  \nonumber \\
\end{eqnarray}
Rearranging
%\begin{eqnarray}
%\partial_t \Gamma_{jk}(\dulR,t) &=&\Gamma_{jk}(\dulR,t) \Bigg\{-i\Delta E_{kj}(\dulR)-\sum_\nu \frac{1}{\vert\chi\vert^2}\bigg[\bigg(i\Delta\nabla_\nu\gamma_{kj}\cdot\bold{J}_\nu+\nabla_\nu\cdot\bold{J}_\nu-\left(\mathfrak{S}_{\nu,k}+\mathfrak{S}_{\nu,j}\right)\cdot\bold{J}_\nu\bigg) \nonumber \\
%&&+\sum_\nu \frac{1}{2M_\nu}\bigg[-i\Delta\mathfrak{Q}_{\nu,jk}-2\left(\pq_{\nu,j}+\pq_{\nu,k}\right)\sum_l\vert C_l\vert^2\nabla_\nu\gamma_{l}+2\left(\pq_{\nu,j}\nabla_\nu\gamma_j+\pq_{\nu,k}\nabla_\nu\gamma_k\right)    \nonumber \\
%&&+2\sum_l\vert C_l\vert^2\left(\nabla_\nu^2\gamma_l-2\mathfrak{S}_{\nu,l}\nabla_\nu\gamma_l\right)-\nabla_\nu^2\left(\gamma_{k}+\gamma_{j}\right)-i(\nabla_\nu\gamma_k)^2+i(\nabla_\nu\gamma_j)^2 \nonumber \\
%&&+2i\Delta\nabla_\nu\gamma_{kj}\sum_l\vert C_l\vert^2\nabla_\nu\gamma_{l}   \Bigg]\Bigg\} \nonumber \\
%\end{eqnarray}
and defining $\nabla_\nu\gamma_k=\mathbf{f}_{\nu,k}$
%\begin{eqnarray}
%\partial_t \Gamma_{jk}(\dulR,t) &=&-\Gamma_{jk}(\dulR,t) \Bigg\{i\Delta E_{kj}(\dulR)+\sum_\nu \frac{1}{\vert\chi\vert^2}\bigg[\bigg(i\Delta\mathbf{f}_{kj}\cdot\bold{J}_\nu+\nabla_\nu\cdot\bold{J}_\nu-\left(\mathfrak{S}_{\nu,k}+\mathfrak{S}_{\nu,j}\right)\cdot\bold{J}_\nu\bigg) \nonumber \\
%&&+\sum_\nu \frac{1}{2M_\nu}\bigg[i\Delta\mathfrak{Q}_{\nu,jk}+2\left(\pq_{\nu,j}+\pq_{\nu,k}\right)\sum_l\vert C_l\vert^2\mathbf{f}_{\nu,l}-2\left(\pq_{\nu,j}\mathbf{f}_{\nu,j}+\pq_{\nu,k}\mathbf{f}_{\nu,k}\right)    \nonumber \\
%&&-2\sum_l\vert C_l\vert^2\left(\nabla_\nu\mathbf{f}_{\nu,l}-2\mathfrak{S}_{\nu,l}\mathbf{f}_{\nu,l}\right)+\nabla_\nu\left(\mathbf{f}_{\nu,k}+\mathbf{f}_{\nu,j}\right)+i\mathbf{f}_{\nu,k}^2-i\mathbf{f}_{\nu,j}^2 \nonumber \\
%&&-2i\Delta\mathbf{f}_{\nu,kj}\sum_l\vert C_l\vert^2\mathbf{f}_{\nu,l}   \Bigg]\Bigg\} \nonumber \\
%\end{eqnarray}
we obtain
\begin{eqnarray}
\partial_t \Gamma_{jk}(\dulR,t) &=&-\Gamma_{jk}(\dulR,t) \Bigg\{i\Delta E_{kj}(\dulR)+\sum_\nu \frac{1}{\vert\chi\vert^2}\bigg[\bigg(i\Delta\mathbf{f}_{kj}\cdot\bold{J}_\nu+\nabla_\nu\cdot\bold{J}_\nu-\left(\mathfrak{S}_{\nu,k}+\mathfrak{S}_{\nu,j}\right)\cdot\bold{J}_\nu\bigg) \nonumber \\
&&+\sum_\nu \frac{1}{2M_\nu}\bigg[i\Delta\mathfrak{Q}_{\nu,jk}+\sum_l\vert C_l\vert^2\left(2\pq_{\nu,j}-\nabla_\nu\right)\cdot\Delta\mathbf{f}_{\nu,lj}+\sum_l\vert C_l\vert^2\left(2\pq_{\nu,k}-\nabla_\nu\right)\cdot\Delta\mathbf{f}_{\nu,lk}\nonumber \\
&&+4\sum_l\vert C_l\vert^2\mathfrak{S}_{\nu,l}\cdot\mathbf{f}_{\nu,l}+i\mathbf{f}_{\nu,k}^2-i\mathbf{f}_{\nu,j}^2-2i\Delta\mathbf{f}_{\nu,kj}\cdot\sum_l\vert C_l\vert^2\mathbf{f}_{\nu,l}   \Bigg]\Bigg\} 
\label{eq:dGammaRtdt}
\end{eqnarray}

%\begin{eqnarray}
%\partial_t \Gamma_{jk}(\dulR,t) &=&-\Gamma_{jk}\Bigg\{\sum_\nu \frac{1}{\vert\chi\vert^2}\bigg(\nabla_\nu\cdot\bold{J}_\nu-\left(\mathfrak{S}_{\nu,k}+\mathfrak{S}_{\nu,j}\right)\cdot\bold{J}_\nu\bigg)+\sum_\nu \frac{1}{2M_\nu}\sum_l\vert C_l\vert^2\nonumber \\
%&&\bigg[\left(2\pq_{\nu,j}-\nabla_\nu\right)\Delta\mathbf{f}_{\nu,lj}+\left(2\pq_{\nu,k}-\nabla_\nu\right)\Delta\mathbf{f}_{\nu,lk}+4\mathfrak{S}_{\nu,l}\mathbf{f}_{\nu,l}\bigg]\nonumber \\
%&&+i\Delta E_{kj}(\dulR)+\sum_\nu \frac{i\Delta\mathbf{f}_{kj}\cdot\bold{J}_\nu}{\vert\chi\vert^2}+\sum_\nu \frac{1}{2M_\nu}\bigg[i\Delta\mathfrak{Q}_{\nu,jk}+i\Delta(\mathbf{f}_{\nu,kj}^2)-2i\Delta\mathbf{f}_{\nu,kj}\sum_l\vert C_l\vert^2\mathbf{f}_{\nu,l}\bigg] \Bigg\} \nonumber \\
%\end{eqnarray}
(where again $\Delta\mathbf{f}_{\nu,lk}=\mathbf{f}_{\nu,l}-\mathbf{f}_{\nu,k}$). 
%Now if we define $\Gamma_{jk}(\dulR,t)=e^{i\int \alpha_{jk}(\dulR,t) dt}\vert\Gamma_{jk}(\dulR,t)\vert$ with the phase of the coherences 
We define $\widetilde\Gamma_{jk} =\Gamma_{jk}e^{i\int^t\alpha_{jk}(\br,t')dt'}$ where 
\begin{equation}
\alpha_{jk}(\dulR,t)=\Delta E_{kj}+\sum_\nu \frac{\Delta\mathbf{f}_{kj}\cdot\bold{J}_\nu}{\vert\chi\vert^2}+\sum_\nu \frac{1}{2M_\nu}\bigg[\Delta\mathfrak{Q}_{\nu,jk}+\Delta(\mathbf{f}_{\nu,kj}^2)-2\Delta\mathbf{f}_{\nu,kj}\cdot\sum_l\vert C_l\vert^2\mathbf{f}_{\nu,l}\bigg]    
\end{equation} 
such that 
\begin{eqnarray}
\partial_t \widetilde\Gamma_{jk}(\dulR,t) =-\widetilde\Gamma_{jk} \Bigg\{\sum_\nu \frac{\big(\nabla_\nu-2\overline{\mathfrak{S}}_{\nu,jk}\big)\cdot\bold{J}_\nu}{\vert\chi\vert^2}+\sum_{\nu,l} \frac{\vert C_l\vert^2}{2M_\nu}\bigg[\overline{\left(4\pq_{\nu,jk}-2\nabla_\nu\right)\cdot\left(\mathbf{f}_{\nu,l}-\mathbf{f}_{\nu,jk}\right)}+4\mathfrak{S}_{\nu,l}\cdot\mathbf{f}_{\nu,l}\bigg]\Bigg\} 
\label{eq:Gammatilde}
\end{eqnarray}
where we use the notation $\overline{g_{jk}} = (g_j + g_k)/2$ to indicate an average over electronic states $j$ and $k$.

 We observe that $ \frac{d\vert\widetilde\Gamma_{jk}\vert^2}{dt}  = 2 \vert\widetilde\Gamma_{jk}\vert \frac{d\vert\widetilde\Gamma_{jk}\vert}{dt}  = 2 {\rm Re} \left(\widetilde\Gamma_{jk}^* \frac{d \widetilde\Gamma_{jk}}{dt}  \right) = -2\vert \widetilde \Gamma_{jk}\vert^2\{..\}$ where $\{..\}$ indicates the term in the curly brackets of Eq.~(\ref{eq:Gammatilde}), which is real. Then using the fact that the magnitudes of $\Gamma$ and $\widetilde\Gamma$ are the same, $ \frac{d\vert\Gamma_{jk}\vert}{dt} = \frac{d\vert\widetilde\Gamma_{jk}\vert}{dt} $, 
%and comparing the second equality in this last equation to the last, 
we find $ \frac{d\vert\Gamma_{jk}\vert}{dt}  = -\vert\widetilde\Gamma_{jk}\vert\{..\} = -\vert\Gamma_{jk}\vert\{..\} $, 
%Then, because $\frac{\partial |\Gamma_{jk}|}{\partial t} = 2|\Gamma_{jk}|^2 {\rm Re}\left(\frac{1}{\widetilde\Gamma_{jk}}\frac{\partial |\widetilde\Gamma_{jk}|}{\partial t}\right)$, and everything in the curly bracket is real, it follows that Eq.~\ref{eq:Gammatilde} holds for $\widetilde\Gamma_{jk}$ replaced by its magnitude, $\vert\Gamma_{jk}\vert$,
as written in Eq. (4) of the main text. 
We stress that this is the exact equation for the magnitude of the coherence in the limit of negligible non-adiabatic couplings.

As in the main text, the trajectory-based equation for the populations and coherences follows from these equations by replacing the nuclear density by a sum over delta-functions centered at each member $I$ of the trajectory ensemble 
\bea
\vert\chi(\dulR,t)\vert^2  &\rightarrow& \sum_I^{N_{tr}}(\vert\chi\vert^2)^{(I)} = \frac{1}{N_{tr}}\sum_I^{N_{tr}}\delta(\dulR - \dulR^{(I)}(t)) \;\;\;\; {\rm and}\\
\Gamma_{jk}(\dulR,t)  &=& \vert \chi(\dulR,t)\vert^2 C_j^*(\dulR,t)C_k(\dulR,t) \rightarrow \frac{1}{N_{tr}}\sum_I^{N_{tr}}\delta(\dulR - \dulR^{(I)}(t))C_j^{*(I)}(t) C_k^{(I)}(t)
\label{eq:trajexp}
\eea
and integrating over $\dulR$. The nuclear velocity and the divergence of the current-density become 
\ben
\frac{{\bf J}_\nu}{\vert\chi\vert^2}\bigg|_{\dulR=\dulR^{(I)}}=\dot\bR_\nu\bigg|_{\dulR=\dulR^{(I)}} \rightarrow \dot\bR^{(I)}_\nu \;\;\;\;\;\;{\rm and}\;\;\;\;\;\;\frac{\nabla_\nu\cdot{\bf J}_\nu}{\vert\chi\vert^2}\bigg|_{\dulR=\dulR^{(I)}}=\partial_t \mathrm{Log}(\vert\chi\vert^2)\bigg|_{\dulR=\dulR^{(I)}} \rightarrow -2 {\bf Q}^{(I)}_\nu\cdot \dot\bR_\nu^{(I)} \;\;\;
\label{eq:Jsimp}
\een
where we have invoked the equation of continuity.
The trajectory's velocity $\dot\bR^{(I)}_\nu$ is given by classical equations of motion derived from the classical nuclear Hamiltonian. Hence, we obtain after integrating Eq.~\ref{eq:dGammaRtdt} over $\dulR$
\begin{eqnarray}
\partial_t \Gamma_{jk}(t) &=&-\sum_I^{N_{tr}}C_{j}^{(I)\,*}C_{k}^{(I)\,*} \Bigg\{i\Delta E^{(I)}_{kj}+\sum_\nu \bigg(i\Delta\mathbf{f}_{kj}^{(I)}\cdot\dot\bR^{(I)}_\nu-2{\bf Q}^{(I)}_\nu\cdot\dot\bR^{(I)}_\nu-\left(\mathfrak{S}_{\nu,k}^{(I)}+\mathfrak{S}_{\nu,j}^{(I)}\right)\cdot\dot\bR^{(I)}_\nu\bigg) \nonumber \\
&&+\sum_\nu \frac{1}{2M_\nu}\bigg[i\Delta\mathfrak{Q}_{\nu,jk}^{(I)}+\sum_l\vert C_l^{(I)}\vert^2\left(2\pq_{\nu,j}^{(I)}-\nabla_\nu\right)\cdot\Delta\mathbf{f}_{\nu,lj}^{(I)}+\sum_l\vert C_l^{(I)}\vert^2\left(2\pq_{\nu,k}^{(I)}-\nabla_\nu\right)\cdot\Delta\mathbf{f}_{\nu,lk}^{(I)}\nonumber \\
&&+4\sum_l\vert C_l\vert^2\mathfrak{S}_{\nu,l}^{(I)}\cdot\mathbf{f}_{\nu,l}^{(I)}+i\mathbf{f}_{\nu,k}^{(I)\,2}-i\mathbf{f}_{\nu,j}^{(I)\,2}-2i\Delta\mathbf{f}_{\nu,kj}^{(I)}\cdot\sum_l\vert C_l^{(I)}\vert^2\mathbf{f}_{\nu,l}^{(I)}   \Bigg]\Bigg\} 
\end{eqnarray}
Further, using the fact that $\overline{\pq}_{\nu,kj}={\bf Q}^{(I)}_\nu-\frac{1}{2}(\mathfrak{S}_{\nu,k}^{(I)}+\mathfrak{S}_{\nu,j}^{(I)})$, we write
\begin{eqnarray}
\partial_t \Gamma_{jk}(t) &=&-\sum_I^{N_{tr}}C_{j}^{(I)\,*}C_{k}^{(I)\,*} \Bigg\{i\Delta E^{(I)}_{kj}+\sum_\nu \bigg(i\Delta\mathbf{f}_{kj}^{(I)}-2\overline{\pq}_{\nu,kj}\bigg)\cdot\dot\bR^{(I)}_\nu+\sum_\nu \frac{1}{2M_\nu}\bigg[i\Delta\mathfrak{Q}_{\nu,jk}^{(I)} \nonumber \\
&&+\sum_l\vert C_l^{(I)}\vert^2\left(2\pq_{\nu,j}^{(I)}-\nabla_\nu\right)\cdot\Delta\mathbf{f}_{\nu,lj}^{(I)}+\sum_l\vert C_l^{(I)}\vert^2\left(2\pq_{\nu,k}^{(I)}-\nabla_\nu\right)\cdot\Delta\mathbf{f}_{\nu,lk}^{(I)}\nonumber \\
&&+4\sum_l\vert C_l\vert^2\mathfrak{S}_{\nu,l}^{(I)}\cdot\mathbf{f}_{\nu,l}^{(I)}+i\mathbf{f}_{\nu,k}^{(I)\,2}-i\mathbf{f}_{\nu,j}^{(I)\,2}-2i\Delta\mathbf{f}_{\nu,kj}^{(I)}\cdot\sum_l\vert C_l^{(I)}\vert^2\mathbf{f}_{\nu,l}^{(I)}   \Bigg]\Bigg\} 
\label{eq:dGamdttraj}
\end{eqnarray} 
We consider the total time-derivative along the trajectory, defined using the convective derivative (Lagrangian frame) $d/dt = \partial_t + \sum_\mu \dot\bR^{(I)}_\nu\cdot\nabla_\nu$. This yields:
\ben
\frac{d\Gamma_{jk}}{dt} \rightarrow \left(\partial_t  + \sum_\nu {\dot\bR}_\nu\cdot\nabla_\nu\right)\Gamma_{jk}  = \partial_t \Gamma_{jk}(t) + \sum_\nu {\dot\bR}_\nu\cdot(i \Delta \mathbf{f}_{\nu,kj} - 2\overline{\pq}_{\nu,kj}) \Gamma_{jk} 
\label{eq:convder}
\een
%\ben
%\frac{d\Gamma_{jk}}{dt} \rightarrow \left(\partial_t + \sum_\nu {\dot\bR}_\nu\cdot\nabla_\nu\right)\Gamma_{jk}  = \left(\partial_t  + \sum_\nu {\dot\bR}_\nu\cdot(i \Delta \nabla_\nu \gamma_{kj} - (\pq_k + \pq_j))\right)\Gamma_{jk} 
%\label{eq:convder}
%\een
Eq.~(\ref{eq:dGamdttraj}) with~(\ref{eq:convder}) gives the {\it exact} equation for the coherences for trajectory-based methods in regions of negligible NAC. Writing in terms of the coefficients (Eq.~(\ref{eq:trajexp})), we obtain
%\begin{equation}
% \sum_I\frac{d(C_j^*C_k)}{dt}^{(I)}=\sum_I\left(\partial_t(C_j^*C_k)^{(I)}-(C_j^*C_k)^{(I)}\sum_\nu\mathbf{\dot{R}}_\nu^{(I)}\cdot\left[i\Delta\mathbf{f}_{\nu,jk}^{(I)}+2\overline{\mathfrak{S}}_{\nu,jk}^{(I)}\right]\right) 
%\end{equation}
%And we write 
\begin{eqnarray}
\sum_I\frac{d(C_j^*C_k)}{dt}^{(I)} &=&-\sum_I C_j^{*(I)}C_k^{(I)} \Bigg\{\sum_{\nu,l} \frac{\vert C_l\vert^2}{2M_\nu}\bigg[\overline{\left(4\pq_{\nu,jk}^{(I)}-2\nabla_\nu\right)\cdot\left(\mathbf{f}_{\nu,l}^{(I)}-\mathbf{f}_{\nu,jk}^{(I)}\right)}+4\mathfrak{S}_{\nu,l}^{(I)}\cdot\mathbf{f}_{\nu,l}^{(I)}\bigg] \nonumber \\
&&+i\Delta E_{kj}^{(I)}+\sum_\nu \frac{1}{2M_\nu}\bigg[i\Delta\mathfrak{Q}_{\nu,jk}^{(I)}+i\Delta(\mathbf{f}_{\nu,kj}^{(I)\,2})-2i\Delta\mathbf{f}_{\nu,kj}^{(I)}\cdot\sum_l\vert C_l^{(I)}\vert^2\mathbf{f}_{\nu,l}^{(I)}\bigg] \Bigg\} \label{eq:dCjCkdt}
\end{eqnarray}
Defining
$\widetilde{{C_j^{(I)*}C_k^{(I)}}}={C_j^{(I)*}C_k^{(I)}}e^{i\int (\alpha_{jk}-\sum_\nu\frac{\Delta\mathbf{f}_{kj}\cdot\mathbf{\dot{R}_\nu}}{\vert\chi\vert^2})dt}$ we can write
\begin{eqnarray}
\frac{d(\widetilde{C_j^*C_k})}{dt}^{(I)} &=&-\widetilde{C_j^{*(I)}C_k^{(I)}} \Bigg\{\sum_\nu \frac{1}{2M_\nu}\sum_l\vert C_l^{(I)}\vert^2\bigg[\overline{\left(4\pq_{\nu,jk}^{(I)}-2\nabla_\nu\right)\cdot\left(\mathbf{f}_{\nu,l}^{(I)}-\mathbf{f}_{\nu,jk}^{(I)}\right)}+4\mathfrak{S}_{\nu,l}^{(\alpha)}\cdot\mathbf{f}_{\nu,l}^{(I)}\bigg]
\end{eqnarray}
%For populations we obtain 
%\begin{eqnarray}
%\frac{d\vert C_k^{(I)}\vert^2}{dt} &=&-\vert %C_k^{(I)}\vert^2 \Bigg\{\sum_\nu \frac{1}{2M_\nu}\sum_l\vert C_l^{(I)}\vert^2\bigg[\left(4\pq_{\nu,k}^{(I)}-2\nabla_\nu\right)\Delta\mathbf{f}_{\nu,lk}^{(I)}+4\mathfrak{S}_{\nu,l}^{(I)}\mathbf{f}_{\nu,l}^{(I)}\bigg]
%\end{eqnarray}
%Which means 
%\begin{eqnarray}
%\left(\frac{d\vert C_k^{(I)}\vert^2}{dt}\right) &=&\left(\frac{d\vert C_k^{(I)}\vert^2}{dt}\right)_{\mathrm{CTMQC}}-\sum_l\vert C_l^{(I)}\vert^2\vert C_k^{(I)}\vert^2 \Bigg\{\sum_\nu \frac{1}{2M_\nu}\bigg[\left(4\mathfrak{S}_{\nu,k}^{(I)}-2\nabla_\nu\right)\Delta\mathbf{f}_{\nu,lk}^{(I)}+4\mathfrak{S}_{\nu,l}^{(I)}\mathbf{f}_{\nu,l}^{(I)}\bigg]\Bigg\}
%\end{eqnarray}
%Now let us check the situations when $j=k$ then we get for the populations
%\begin{eqnarray}
%\partial_t \vert\chi_{k}(\dulR,t)\vert^2 &=&-%\vert\chi_{k}(\dulR,t)\vert^2\sum_\nu \Bigg\{ \frac{1}{\vert\chi\vert^2}\bigg[\nabla_\nu\cdot\bold{J}_\nu-2\mathfrak{S}_{\nu,k}\cdot\bold{J}_\nu\bigg]+\frac{1}{M_\nu}\bigg[\sum_l\vert C_l\vert^2(2\pq_{\nu,k}+\nabla_\nu)\cdot(\mathbf{f}_{\nu,l}-\mathbf{f}_{\nu,k})    \nonumber \\
%&&+2\sum_l\vert C_l\vert^2\mathfrak{S}_{\nu,l}\mathbf{f}_{\nu,l}\bigg]\Bigg\}
%\end{eqnarray}
which by the same argument as given below Eq.~\ref{eq:Gammatilde} the equation holds for $\vert C_j^*C_k\vert^{(I)}$, as given in Eq. 10 of the main text. 

Finally, to define the exact trajectory-based propagation equations, we now go back and include the terms from the NAC.
\ben
\partial_t \Gamma_{jk}(\dulR,t) = \partial_t \Gamma_{jk}(\dulR,t)^{\rm Eq.~\ref{eq:dGammaRtdt}} + \partial_t \Gamma_{jk}(\dulR,t)^{\rm NAC}
\een
where
\begin{eqnarray}
&&\partial_t \Gamma_{jk}(\dulR,t)^{\rm NAC}=-\sum_{\nu,n\neq k}\frac{\Gamma_{jn}}{2M_\nu}\bigg[-iD_{kn,\nu}+2\bold{d}_{kn,\nu}\cdot\left(\frac{M_\nu\mathbf{J}_\nu}{\vert\chi\vert^2}+\sum_l\vert C_l\vert^2\Delta\mathbf{f}_{\nu,nl}+i \sum_{ml} C_l^*C_m {\bf d}_{\nu,lm}+i\pq_{\nu,n})\right)\bigg]\nonumber \\
&&+\sum_{\nu,n\neq j}\frac{\Gamma_{nk}}{2M_\nu}\bigg[-iD_{nj,\nu}-2\bold{d}_{jn,\nu}\cdot\left(\frac{M_\nu\mathbf{J}_\nu}{\vert\chi\vert^2}+\sum_l\vert C_l\vert^2\Delta\mathbf{f}_{\nu,nl}+i\sum_{ml} C_l^*C_m {\bf d}_{\nu,lm}-i\pq_{\nu,n}\right)\bigg]\nonumber \\
&&+\sum_\nu\frac{\Gamma_{jk}}{2M_\nu}\bigg[4i\overline{\pq}_{\nu,jk}\cdot\sum_{lm}C_l^*C_m\mathbf{d}_{\nu,lm}+2\sum_{lm}\left(\Delta\mathbf{f}_{\nu,lm}-2i\overline{\mathfrak{S}}_{\nu,lm}+i\nabla_\nu\right)\cdot\mathbf{d}_{\nu,lm} -2\Delta\mathbf{f}_{\nu,jk}\cdot\sum_{lm}C_l^*C_m\mathbf{d}_{\nu,lm} \bigg]
\end{eqnarray}
so that for the trajectory-based coherences and populations, we obtain
\begin{equation}
\sum_I\frac{d (C^{*(I)}_jC_k^{(I)})}{dt} = \sum_I \left(\frac{dC^{*(I)}_jC_k^{(I)}}{dt} \right)^{Eq.~\ref{eq:dCjCkdt}} + \left(\frac{dC^{*(I)}_jC_k^{(I)}}{dt} \right)^{\rm NAC} 
\end{equation}
where
\begin{eqnarray}
&&\left(\frac{dC^{*(I)}_jC_k^{(I)}}{dt} \right)^{\rm NAC}=-\sum_{\nu,n\neq k}\frac{(C_j^*C_n)^{(I)}}{2M_\nu}\bigg[-iD_{kn,\nu}^{(I)}+2\bold{d}_{kn,\nu}^{(I)}\cdot\bigg(M_\nu\mathbf{\dot{R}}_\nu^{(I)}+\sum_l\vert C_l^{(I)}\vert^2\Delta\mathbf{f}_{\nu,nl}^{(I)}+i \sum_{ml} (C_l^*C_m)^{(I)} {\bf d}_{\nu,lm}^{(I)}\nonumber \\
&&+i\pq_{\nu,n})^{(I)}\bigg)\bigg]+\sum_{\nu,n\neq j}\frac{(C_n^*C_k)^{(I)}}{2M_\nu}\bigg[-iD_{nj,\nu}^{(I)}-2\bold{d}_{jn,\nu}^{(I)}\cdot\left(M_\nu\mathbf{\dot{R}}_\nu^{(I)}+\sum_l\vert C_l^{(I)}\vert^2\Delta\mathbf{f}_{\nu,nl}^{(I)}+i\sum_{ml} (C_l^*C_m)^{(I)} {\bf d}_{\nu,lm}^{(I)}-i\pq_{\nu,n}^{(I)}\right)\bigg]\nonumber \\
&&+\sum_\nu\frac{(C_j^*C_k)^{(I)}}{M_\nu}\sum_{lm}C_l^{*(I)}C_m^{(I)}\bigg[2i\overline{\pq}_{\nu,jk}^{(I)}+\Delta\mathbf{f}_{\nu,lm}^{(I)}-2i\overline{\mathfrak{S}}_{\nu,lm}^{(I)}+i\nabla_\nu+\Delta\mathbf{f}_{\nu,kj}^{(I)} \bigg]\cdot\mathbf{d}_{\nu,lm}^{(I)} \nonumber \\
\end{eqnarray}

\section{SM.3 Simulation details}
For the quantum dynamics (QD) simulations, the time-dependent Schrödinger equation is solved on a grid in the diabatic basis using the split-operator method. The system is initialized in a coherent superposition of three gaussian nuclear wavepackets
\begin{equation}
\vert\Psi (R,0)\rangle=\chi(R,0)\left(\vert\phi_{R,1}\rangle+\vert\phi_{R,2}\rangle/2+\vert\phi_{R,3}\rangle/2    \right)
\label{eq:superpos}
\end{equation}
with $\chi(R,0)=\left(\frac{2\pi^{1/2}}{3\sigma}\right)^{1/2}e^{-\frac{(R-R_0)^2}{2\sigma^2}+i(R-R_0)P_0}$ . The initial variance, position, momentum are $(\sigma,R_0,P_0)=(1\,\mathrm{bohr},-26\,\mathrm{bohr}^{-1},40\,\mathrm{bohr})$ for the EL20-SAC model and $(\sigma,R_0,P_0)=(0.223\,\mathrm{bohr},-4\,\mathrm{bohr},0\,\mathrm{bohr}^{-1})$ for the 3HO model. For the EL20-SAC model the spatial grid is defined in the range $[-26:28]$ bohr with 4000 grid points and with a time-step of $dt=0.0012$ fs. For the 3HO model the chosen spatial grid range is $[-10:10]$ bohr with $2000$ grid points and with a time-step of $dt=0.0024$ fs.

The trajectory-based simulations were performed in the \textsc{G-CTMQC} package~\cite{GCTMQC} and the SHXF simulations in the \textsc{PyUNIXMD} package~\cite{PyUNIxMD}. In both models 1000 Wigner-sampled trajectories were run using the same initial conditions as for the exact case. The initial trajectory ensemble for all methods, was initialized in a pure ensemble, i.e each trajectory carries the same coherent superposition of electronic
states. The active states for the surface hopping simulations were stochastically selected to match the net populations at
the initial time. We refer the reader to Ref.~\cite{VMA24} for a study on different choices (pure and mixed) of initial electronic state in trajectory-based methods.

\subsection{Example 1: EL20-SAC}
The Hamiltonian in the diabatic basis is given by
\begin{equation}
H(R)= \begin{pmatrix}
V_1(R) & 0 & 0 \\
0 & V_2(R) & \lambda (R) \\
0 & \lambda (R) & -V_2(R)
\end{pmatrix}  
\end{equation}
with
\begin{eqnarray}
V_{1}(R)&=&-0.03(R+35)-0.02 \nonumber \\
V_{2}(R)&=&\frac{R}{\vert R\vert} A \left(1-e^{-B\vert R\vert}\right)\nonumber \\
\lambda(R)&=&Ce^{-DR^2}
\end{eqnarray}
the chosen model parameters were $A=0.01\,\mathrm{Ha}$, $B=1.6\,\mathrm{bohr}^{-1}$ $C=0.005\,\mathrm{Ha}$ and $D=1\,\mathrm{bohr}^{-2}$.

\subsection{Example 2: 3HO}
The  Hamiltonian reads
\begin{equation}
H(R)= \begin{pmatrix}
\frac{1}{2}k_1R^2 & 0 & 0 \\
0 & \frac{1}{2}k_2R^2+\Delta & 0 \\
0 & 0 & \frac{1}{2}k_2R^2+2\Delta 
\end{pmatrix}  
\end{equation}
with $k_1=0.005 \mathrm{Ha}^2$, $k_2=0.02 \mathrm{Ha}^2$ and $\Delta=0.01\,\mathrm{Ha}$.
%with $k_1=0.005 \mathrm{Ha}^2m_e/\hbar^2$, $k_2=0.02 \mathrm{Ha}^2m_e/\hbar^2$ and $\Delta=0.01\,\mathrm{Ha}$.

\section{SM.4 Effect of $\mathfrak{S}$}
To show the importance of $\mathfrak{S}(R,t)$ we plotted time-snapshots of the exact $Q_1(R,t)$, $\mathfrak{S}_1(R,t)$ and $\mathcal{Q}_1(R,t)$ during the first recoherence event. We observe both the nuclear quantum momentum $Q_1$ and the reduced contribution $\mathfrak{S}_1$ are active during this event, and the large features that $Q_1$ displays are offset largely by $\mathfrak{S}_1$ yielding a relatively small $\mathcal{Q}_1$. As discussed in the main text, the $\mathfrak{S}$ terms are responsible to induce recoherence away from NAC regions causing a change in the electronic coherences which then activates the terms dependent on $Q(R,t)$.
\begin{figure}[h]
    \centering
\includegraphics[width=0.8\textwidth]{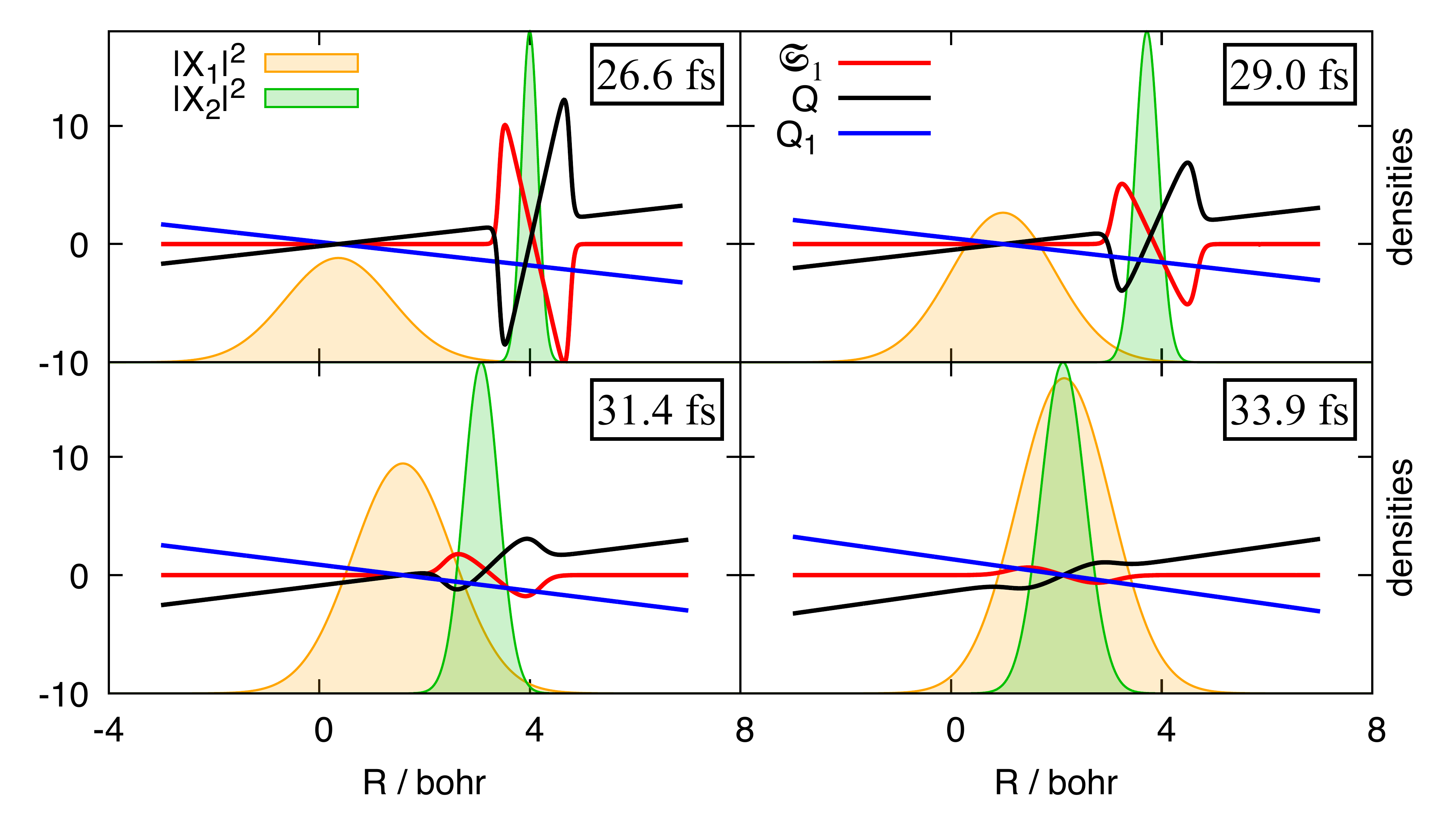}
    \caption{Time snapshots of the exact density on the first (orange) and second (green) electronic states, nuclear quantum momentum (blue line), and projected quantum momentum (black line) and crunch term (red line) on state 1 during the recoherence event for 3HO model.}
    \label{fig:crunch}
\end{figure}

%\bibliography{ref_SI,ref_na}

\end{document}